\documentclass[letter, 12pt]{article}

\usepackage[margin=1in]{geometry}
\usepackage[numbers,sort&compress]{natbib} 
\usepackage{authblk} 
\usepackage{color}
\usepackage{amsmath}
\usepackage{amssymb} 
\usepackage{graphicx} 
\usepackage{fancyhdr} 
\usepackage{amsfonts}
\usepackage{tabularx} 
\usepackage{slashed}
\usepackage[
singlelinecheck=false 
]{caption}

\def\mpl{\text{M}_\text{pl}}
\def\Sb{\bar{S}}
\def\be{\begin{equation}}
\def\ee{\end{equation}}
\def\bea{\begin{eqnarray}}
\def\eea{\end{eqnarray}}
\def\half{\frac{1}{2}}
\newcommand{\lrs}[1]{\left[{#1}\right]}
\newcommand{\lrp}[1]{\left({#1}\right)}
\newcommand\zfr{$\mathbb{Z}_4^R$~}
\newcommand\zfrmm{\mathbb{Z}_4^R~}
\newcommand{\fw}{\ensuremath{\mathcal{W}}}
\newcommand{\fk}{\ensuremath{\mathcal{K}}}
\newcommand{\im}[1]{\text{Im}{(#1)}}
\newcommand{\fo}{\ensuremath{\mathcal{O}}}

\begin{document}

\renewcommand\headrule{} 

\title{\huge \bf{Reheating and Leptogenesis after Pati-Salam F-term Subcritical Hybrid Inflation}}
\author{B. Charles Bryant}
\author{Zijie Poh}
\author{Stuart Raby}
\affil{\emph{Department of Physics}\\\emph{The Ohio State University}\\\emph{191 W.~Woodruff Ave, Columbus, OH 43210, USA}}

\maketitle
\thispagestyle{fancy}
\pagenumbering{gobble} 

\begin{abstract}\normalsize\parindent 0pt\parskip 5pt

In this paper, we extend the analysis of a Pati-Salam subcritical hybrid $F$-term inflation model, proposed by two of us~\cite{Bryant:2016tzg}, by studying the reheating and the baryogenesis  (via leptogenesis) of the model.  This SUSY GUT model is able to fit low energy electroweak precision data, LHC data, $b$-physics data, in addition to inflation observables such as the tensor-to-scalar ratio and the scalar spectral index.  The reheating mechanism of this model is instant preheating due to the bosonic and fermionic broad parametric resonance, while the baryon-to-entropy ratio is obtained from the CP asymmetric right-handed (s)neutrinos decay.  The phases in the neutrino Yukawa matrices are fixed by fitting to the low energy observables.  With these phases, the heaviest right-handed (s)neutrinos decay to produce a lepton asymmetry with the correct sign, while the two lighter right-handed (s)neutrinos decay to produce the wrong sign.  Consequently, the baryogenesis analysis is necessarily performed by including all three families of the right-handed (s)neutrinos.

\end{abstract}

\pagenumbering{arabic} 

\newpage
\section{Introduction}
One path to Grand Unified Theories (GUTs) without large GUT representations is via 5D or 6D orbifold GUTs.  One possible 4D gauge symmetry resulting from orbifolding a higher dimensional GUT is Pati-Salam (PS) gauge symmetry, $\text{SU}(4)_\text{C}\times\text{SU}(2)_\text{L}\times\text{SU}(2)_\text{R}$~\cite{Kobayashi:2004ud,Kobayashi:2004ya}.  Due to the higher dimensional GUT completion, the gauge couplings of the theory are unified.  Recently, two of us have introduced a $\mathbb{Z}_4^R$ discrete symmetry to the PS gauge symmetry to obtain a model of inflation, with so-called subcritical $F$-term hybrid inflation, which fits the tensor-to-scalar ratio, the scalar spectral index, and the scalar power spectrum~\cite{Bryant:2016tzg}\footnote{It has been shown that the $\mathbb{Z}_4^R$ symmetry forbids the SUSY $\mu$ term and dimension 4 and 5 proton decay operators to all orders in perturbation theory~\cite{Lee:2010gv,Lee:2011dya}. Moreover, non-perturbative effects can then generate the $\mu$ term and suppress dimension 5 proton decay, while preserving R-parity.}.  Coupled with previous results, which showed that the matter sector of the model can fit the low-energy observables~\cite{Raby:2015rex,Poh:2015wta,Anandakrishnan:2014nea,Raby:2013wva,Anandakrishnan:2013pja,Anandakrishnan:2013nca,Anandakrishnan:2012tj}, this model is a complete theory.  Under PS symmetry$\times\mathbb{Z}_4^R$, matter in one family is unified into two irreducible representations: $Q=(4,2,1,1)$ and $Q^c=(\bar{4},1,\bar{2},1)$, while the Higgs doublets are unified into a single irreducible representation: $\mathcal{H}=(1,2,\bar{2},0)$.  On the other hand, the inflaton sector contains an inflaton field and two waterfall fields: $\Phi=(1,1,1,2), S^c=(\bar{4},1,\bar{2},0)$, and $\bar{S}^c=(4,1,2,0)$, respectively.  In this paper, we provide a detailed discussion of the reheating process and the baryogenesis of the model.

Unlike most leptogenesis analyses in the literature~\cite{Buchmuller:2005eh,Davidson:2008bu,Bjorkeroth:2016lzs}, we do not have the privilege to integrate out the two heavier right-handed (s)neutrinos in our analysis.  By fitting to the low energy observables, the heaviest right-handed (s)neutrinos decay to produce the correct sign for the lepton asymmetry while the two lighter ones decay to produce the the wrong sign.  Another interesting feature of this model is that reheating occurs via the process of instant preheating~\cite{Kofman:1997yn,Felder:1998vq,Greene:2000ew}.  As a consequence of a broad parametric resonance, particle creation occurs in a discrete manner.  A lepton asymmetry is induced when the inflaton creates Higgses non-perturbatively and the Higgses subsequently decay to right-handed (s)neutrinos.  A free parameter, $\alpha$, controls the inflaton-Higgs coupling and thus the amount of the final asymmetry.  Hence, this model is not constrained by the measured baryon-to-entropy ratio.  Instead, it is constrained by low-energy data and inflation observables.

It is important to point out that we are assuming that supersymmetry is broken by gravity mediation, hence gravitinos in our model have mass $m_{3/2}\gtrsim40\,\text{TeV}$~\cite{Kawasaki:2008qe,Gherghetta:1999sw,Weinberg:1982zq,Ellis:1984er,Ellis:1984eq}.  Heavy gravitinos are required to ensure that the gravitino in our model decays before big bang nucleosynthesis; thus causing no cosmological problem.  In addition, we also assume that the cosmological moduli problem is ameliorated by having all moduli with mass at the GUT scale~\cite{Coughlan:1983ci,Ellis:1986zt,Banks:1995dt,Banks:1993en}.

In Sec.~\ref{sec:review}, we briefly review the model of Ref.~\cite{Bryant:2016tzg}.  The mechanism of instant preheating is explained in Sec.~\ref{sec:broad_stroke}, which is based on the detailed discussion of bosonic and fermionic broad parametric resonances found in~\cite{Kofman:1997yn,Felder:1998vq,Greene:2000ew}.  Also included in Sec.~\ref{sec:broad_stroke} are parts of the model relevant to reheating and baryogenesis such as the decay of the inflaton and the waterfall field, and the generation of the lepton asymmetry.  In Sec.~\ref{sec:evolution}, the evolution equations for evolving the system from the end of inflation to the decay of the right-handed (s)neutrinos are presented.  The parameters of the model and the simulation procedure are outlined in Sec.~\ref{sec:procedure}, while the results and discussions are contained in Sec.~\ref{sec:results}.  The analysis of this paper is very similar to the analysis in Ahn and Kolb~\cite{Ahn:2005bg}.

\section{Brief Review of PS model}
\label{sec:review}

In this section, we briefly review the results of Ref.~\cite{Bryant:2016tzg}.  The superpotential and K\"{a}hler potential for the inflaton sector of the model with a Pati-Salam $SU(4)_C \times SU(2)_L \times SU(2)_R$ gauge symmetry times \zfr discrete $R$ symmetry are given by
\begin{align}
  \fw_{I} &= \Phi \left( \kappa \Sb^c S^c + m_\phi Y + \frac{1}{\sqrt{2}}\alpha {\cal H} {\cal H}\right) + \lambda X\lrp{\Sb^c S^c-\frac{v^2_{PS}}{2}} + S^c \Sigma S^c + \Sb^c \Sigma \Sb^c \label{eq:superpotential}\\
  \fk &= \half(\Phi+\Phi^\dagger)^2 + (S^c)^\dagger S^c + (\Sb^c)^\dagger \Sb^c + Y^\dagger Y + X^\dagger X\lrs{1-c_X \frac{X^\dagger X}{\mpl^2}+a_X\lrp{\frac{X^\dagger X}{\mpl^2}}^2 } \,,
\end{align}
with the quantum numbers of the inflaton and waterfall superfields: $ \{\Phi = (1, 1, 1, 2), \; S^c = (\bar{4}, 1, \bar{2}, 0), \; \Sb^c = (4, 1, 2, 0) \} $.  As a consequence, the Pati-Salam gauge symmetry is broken to the Standard Model (SM) at the waterfall transition and remains this way both during inflation and afterwards.  The superfield, $\Sigma = (6,1,1,2)$, is needed to guarantee that the effective low energy theory below the PS breaking scale is just the minimal supersymmetric standard model (MSSM).  The singlet $X = (1,1,1,2)$ is introduced in order to obtain $F$-term hybrid inflation in which the coupling of the inflaton to the waterfall field is independent of the self-coupling of the waterfall field.  The term with the singlet $Y=(1,1,1,0)$ is added in order to obtain a supersymmetric vacuum after inflation.  The parameter $m_\phi \sim 10^{-6} \mpl$ (where $\mpl = 2.4 \times 10^{18}$ GeV is the reduced Planck scale) is smaller than in typical chaotic inflation models and the $F$-term of $Y$ acts to lift the flatness of the potential above the critical point.  The term with the Higgs field, ${\cal H}= (1, 2, \bar 2, 0)$, is added to enable reheating, which will be discussed later.  The K\"{a}hler potential has a shift symmetry, $\im{\Phi} \rightarrow \im{\Phi} + \Theta$, where $\Theta$ is a real constant.  The parameter $\kappa \ll \lambda$ such that the critical value (where the sign of the waterfall field mass squared becomes negative) $\phi_c = \frac{\lambda v_{PS}}{\kappa} \gg \mpl$.  This is the parameter regime where subcritical hybrid inflation occurs~\cite{Buchmuller:2014rfa,Buchmuller:2014dda}.  Since PS is spontaneously broken during inflation there is no monopole problem.  Performing a $\chi^2$ fit to cosmological data, a best fit point was found with  $\kappa \simeq 4.5\times10^{-4}$, $\lambda \simeq0.8$, $m_\phi=10^{-6}\mpl$ and $v_{PS} \simeq 1.25\times10^{-2}\mpl\simeq 3\times 10^{16}\text{GeV}$.  With these parameter values, 60 e-foldings of inflation started at $\phi_*=14.5~\mpl$ and the cosmological observables are computed to be
\be
r = 0.084 \,,
\qquad n_s = 0.963 \,,
\qquad A_s =  2.21\times10^{-9}
\, .\ee

The matter sector of the theory is given by the superpotential $\fw = \fw_I + \fw_{PS}$ with
\begin{align}
  \begin{aligned}
    \fw_{PS} =&\, \fw_{neutrino} + \lambda Q_3 {\cal H} Q^c_3 + Q_a {\cal H} F_a^c + F_a {\cal H} Q_a^c
    \\& + \bar F_a^c \left( M F_a^c + \phi_a {\cal O_{B-L}} Q_3^c + {\cal O_{B-L}} \frac{\theta_a \theta_b}{\hat M} Q_b^c + B_2 Q_a^c\right)
    \\& + \bar F_a \left( M F_a + \phi_a {\cal O_{B-L}} Q_3 + {\cal O_{B-L}} \frac{\theta_a \theta_b}{\hat M} Q_b + B_2 Q_a\right) \,,
  \end{aligned}
\end{align}
where  $ \{ Q_3, \ Q_a, \ F_a \} = (4, 2, 1,1), \;  \{ Q_3^c, \ Q^c_a, \ F_a^c \} = (\bar 4, 1, \bar 2, 1)$ with $a = 1,2$, a $D_4$ family index, ${\cal H} = (1, 2, \bar 2, 0)$ and the fields $\bar F_a,  \ \bar F_a^c$ are Pati-Salam conjugate fields.  The superpotential for the neutrino sector is given by
\begin{align}
  \begin{aligned}
    \fw_{neutrino} &= \Sb^c ( \lambda_2 N_a Q^c_a + \lambda_3 N_3 Q^c_3 )
    - \frac{1}{2} \left( \lambda'_2 Y^\prime N_a N_a + \frac{\tilde \theta_a \tilde \theta_b}{\hat M} N_a N_b + \lambda'_3 Y^\prime N_3 N_3 \right)
    \\&= \sum_{i=1}^3\frac{\lambda_i^2}{2M_i}(\bar{S}^cQ_i^c)^2 \,,
  \end{aligned} \label{eq:Wnu}
\end{align}
where
\be
  M_1 = \lambda'_2 Y^\prime \, \qquad M_2 = \lambda'_2 Y^\prime + \frac{\tilde \theta_2^2}{\hat M}\, \qquad  M_3 = \lambda'_3 Y^\prime \,,
\ee
and $\widetilde{\theta}_1$ is taken to be zero.\footnote{The fields $Y$ and $Y^\prime$ can be distinguished by an additional $\mathbb{Z}_4$ symmetry where $Y$ is invariant, but $Y^\prime, \ N_a, \ N_3, \ \tilde \theta_a, \ S^c, \ \Sb^c, \ \Sigma$ have $\mathbb{Z}_4$ charges $2, \ 1, \ 1, \ 1, \ 1,  \ 3, \ 2$, respectively.}

After expanding the waterfall field by its vacuum expectation value (vev), the last line of eq.~(\ref{eq:Wnu}) yields (with $\Sb^c \rightarrow V^c/\sqrt{2}$)
\be
  \frac{\lambda^2_i}{2 M_i}\lrp{\frac{\sigma+i\tau+\sqrt{2}v_{PS}}{2}}^2\bar \nu_i \bar \nu_i
  = \half M_{R_i} \bar \nu_i \bar \nu_i + \frac{h_i}{2}\lrp{\sigma+i\tau}\bar \nu_i \bar \nu_i \,,
  \label{eq:sdecay}
\ee
plus terms quadratic in $\sigma$ and $\tau$ with
\begin{align}
  M_{R_i}\equiv \frac{\lambda^2_i v^2_{PS}}{2 M_i}
  \hspace{1cm}\text{and}\hspace{1cm}
  h_i \equiv \frac{\lambda^2_i v_{PS}}{\sqrt{2}M_i} \,,
  \label{eq:MRi_hi}
\end{align}
where $\lambda_1 = \lambda_2$.

Here $Y^\prime$ is identified as one of the flavon fields. The ``right-handed" neutrino fields,  $N_a, \ N_3$ are PS singlets with charge $(1,1,1,1)$. The vev of $Y^\prime$ gives a heavy mass term for $N_a, \ N_3$ which are in turn integrated out to yield effective couplings between the waterfall field and the left-handed anti-neutrinos in $Q^c_a$ and $Q^c_3$. Similar to the waterfall field, the scalar components of $Y$ also obtain a coupling to the left-handed anti-neutrinos
\be
\frac{h_i}{2}\lrp{\frac{m}{\kappa v_{PS}}}\lrp{h+iu}\bar \nu_i \bar \nu_i
\,.
\label{eq:hdecay}
\ee

The fields $F_a, \ \bar F_a, \ F_a^c, \ \bar F_a^c$ are Froggatt-Nielson fields which are integrated out to obtain the effective Yukawa matrices.  The effective operators $\cal O_{B-L}$ and $\cal O$ are defined by
\begin{align}
  \begin{aligned}
    {\hat M^2 ({\cal O_{B-L}})^{\alpha i}}_{\beta j} \equiv& - \frac{4}{3} {\delta^i}_j \bar {S^c}^{\gamma k} \lrp{{\delta^\alpha}_\gamma {\delta^\lambda}_\beta - \frac{1}{4} {\delta^\alpha}_\beta {\delta^\lambda}_\gamma}  S^c_{\lambda k}
    \\=& {(B-L)^\alpha}_\beta {\delta^i}_j \frac{v_{PS}^2}{2} \,,
  \end{aligned}
\end{align}
and
\begin{align}
  \begin{aligned}
    {\hat M^2 {\cal O}^{\alpha i}}_{\beta j} &\equiv \bar {S^c}^{\gamma k} \left[{\delta^\alpha}_\beta {\delta^i}_j {\delta^\lambda}_\gamma {\delta^l}_k + \tilde \alpha {\delta^\lambda}_\gamma \lrp{ {\delta^i}_k {\delta^l}_j - \frac{1}{2} {\delta^i}_j {\delta^l}_k}\right.
    \\&\hspace{2.8cm} \left. -\frac{4}{3} \tilde \beta {\delta^l}_k {\delta^i}_j \lrp{{\delta^\alpha}_\gamma {\delta^\lambda}_\beta - \frac{1}{4} {\delta^\alpha}_\beta {\delta^\lambda}_\gamma} \right] S^c_{\lambda l}
    \\&= \lrs{{\mathbb{I}^{\alpha i}}_{\beta j} + \tilde \alpha {(T_{3R})^i}_j {\delta^\alpha}_\beta + \tilde \beta {(B-L)^\alpha}_\beta {\delta^i}_j } \frac{v_{PS}^2}{2}
    \\&\equiv \lrs{{\mathbb{I}^{\alpha i}}_{\beta j} + \alpha {(X)^{i \alpha}}_{j \beta} + \beta {(Y)^{i \alpha}}_{j \beta} } \frac{v_{PS}^2}{2} \,,
  \end{aligned}
\end{align}
where $X = 3 (B - L) - 4 T_{3R}$ commutes with $SU(5)$ and $Y = 2 T_{3R} + (B - L)$ is the SM hypercharge.  The Froggatt-Nielson fields $F_a, \ \bar F_a, \  F_a^c, \ \bar F_a^c$ have a mass term $M$ given by $M_0 \ {{\fo}^{\alpha i}}_{\beta j}$.  The flavon fields $\phi_a, \ \theta_a, \ \tilde \theta_a$ are doublets under $D_4$ while $B_2$ is a non-trivial $D_4$ singlet such that the product $B_2 * (x_1 y_2 - x_2 y_1)$ is $D_4$ invariant with $x_a, \ y_a$ as $D_4$ doublets.  The $D_4$ invariant product between two doublets is given by $x_a y_a \equiv  x_1 y_1 + x_2 y_2$.  All flavon fields have zero charge under \zfr.  The flavon fields $\phi_{1,2}, \ \theta_2, \ \tilde \theta_2, \ B_2$ are assumed to get non-zero vevs while all other flavon fields have zero vevs.

Note, with the given particle spectrum and \zfr charges, we have the following anomaly coefficients, \be A_{SU(4)_C-SU(4)_C-\zfrmm} = A_{SU(2)_L-SU(2)_L-\zfrmm} = A_{SU(2)_R-SU(2)_R-\zfrmm} = 1 ({\rm mod}(2)) . \ee Thus the \zfr anomaly can, in principle, be canceled via the Green-Schwarz mechanism, as discussed in Ref.~\cite{Lee:2010gv,Lee:2011dya}.  Dynamical breaking of the \zfr symmetry would then preserve an exact $R$-parity and generate a $\mu$ term, with $\mu \sim m_{3/2}$ and dimension 5 proton decay operators suppressed by  $m_{3/2}^2/\text{M}_\text{pl}$.

\subsection{Yukawa matrices}
Upon integrating out the heavy Froggatt-Nielsen fields, we obtain the effective superpotential for the low energy theory,
\begin{align}
    \fw_{LE} =  Y^u_{i j} \ q_i \ H_u \ \bar u_j + Y^d_{i j} \ q_i \ H_d \ \bar d_j + Y^e_{i j} \ \ell_i \ H_d \ \bar e_j + Y^\nu_{i j} \ \ell_i \ H_u \ \bar \nu_j + \frac{1}{2} \  M_{R_i} \bar \nu_i \ \bar \nu_i \,,
\end{align}
where $i,j=1,2,3$ and
\be
  M_{R_{1,2}} = \frac{\lambda_2^2 \ v_{PS}^2}{2 \ M_{1,2}} \,,
  \quad
  M_{R_{3}} = \frac{\lambda_3^2 \ v_{PS}^2}{2 \ M_{3}} \,.
  \label{eq:masses}
\ee
The Yukawa matrices for up-quarks, down-quarks, charged leptons and neutrinos are given by (defined in Weyl notation with doublets on the left)\footnote{\label{fn:yukawa} These Yukawa matrices are identical to those obtained previously (see Ref.~\cite{Dermisek:2005ij}) and analyzed most recently in Ref.~\cite{Anandakrishnan:2012tj,Anandakrishnan:2014nea}.}
\begin{align}
  \begin{aligned}
      Y^u =& \left(\begin{array}{ccc} 0                & \epsilon'\ \rho      & -\epsilon\ \xi \\
                                      -\epsilon'\ \rho & \tilde\epsilon\ \rho & -\epsilon      \\
                                      \epsilon\ \xi    & \epsilon             & 1
                   \end{array}\right)\;\lambda
    \\Y^d =& \left(\begin{array}{ccc} 0             & \epsilon'       & -\epsilon\ \xi\ \sigma \\
                                      -\epsilon'    & \tilde\epsilon  & -\epsilon\ \sigma      \\
                                      \epsilon\ \xi & \epsilon        & 1
                    \end{array}\right)\;\lambda
    \\Y^e =&  \left(\begin{array}{ccc} 0                         & -\epsilon'           & 3\ \epsilon \ \xi \\
                                       \epsilon'                 & 3\ \tilde\epsilon    & 3\ \epsilon       \\
                                       -3\ \epsilon\ \xi\ \sigma & -3\ \epsilon\ \sigma & 1
                    \end{array}\right)\;\lambda \,,
  \end{aligned}\label{eq:yukawaD31}
\end{align}
with
\begin{align}
  \begin{aligned}
      \xi      &=       \phi_1/\phi_2\,,             &&&\tilde\epsilon  &\propto (\theta_2/\hat M)^2\,,
    \\\epsilon &\propto -\phi_2/\hat M\,,            &&&\epsilon^\prime &\sim    ({ B_2}/M_0),
    \\\sigma   &=       \frac{1+\alpha}{1-3\alpha}\,,&&&\rho            &\sim    \beta\ll\alpha \,,
  \end{aligned}\label{eq:omegaD3}
\end{align}
and
\bea
  Y^{\nu} = \left(\begin{array}{ccc} 0                         & -\epsilon'\ \omega        & {3 \over 2}\ \epsilon\ \xi\ \omega \\
                                     \epsilon'\ \omega         & 3\ \tilde\epsilon\ \omega & {3 \over 2}\ \epsilon\ \omega      \\
                                     -3\ \epsilon\ \xi\ \sigma & -3\ \epsilon\ \sigma      & 1
                  \end{array} \right) \;
            \lambda
  \label{eq:yukawaD32} \,,
\eea
with $\omega=2\,\sigma/(2\,\sigma-1)$ and a Dirac neutrino mass matrix given by
\begin{equation}
  m_\nu \equiv Y^\nu\frac{v}{\sqrt{2}}\sin\beta \,.
  \label{eq:mnuD3}
\end{equation}

From eq.~(\ref{eq:yukawaD31}) and (\ref{eq:yukawaD32}), one can see that the flavor hierarchies in the Yukawa couplings are encoded in terms of the four complex parameters $\rho, \sigma, \tilde \epsilon, \xi$ and three real parameters $\epsilon, \epsilon', \lambda$.  These matrices contain 7 real parameters and 4 arbitrary phases.  While the superpotential $\fw_{PS}$ has many arbitrary parameters, the resulting effective Yukawa matrices have much fewer parameters, therefore obtaining a very predictive theory.  Also, the quark mass matrices accommodate the Georgi-Jarlskog mechanism, such that $m_\mu/m_e \approx 9 \ m_s/m_d$.  This is a result of the operator $\cal O_{B-L}$ which is assumed to have a vev in the $B - L$ direction.

\section{Instant Preheating in Broad Strokes}
\label{sec:broad_stroke}
After inflation the universe must reheat.  This reheating occurs via the process of instant preheating.  At the same time, an asymmetry in the number of leptons minus anti-leptons can also be obtained.  Let us now describe this process.

In this model, the inflaton superfield, $\Phi$, couples to the Higgses superfield, $\mathcal{H}$, via the following operator:
\begin{align}
  \mathcal{W} &= \frac{1}{\sqrt{2}}\alpha\Phi\mathcal{H}\mathcal{H} \,.
\end{align}
Including the Yukawa term, the superpotential is
\begin{align}
  \mathcal{W}
  = \sqrt{2}\alpha\Phi H_uH_d +
    \lambda_{u;ij}\bar{u}_iH_uq_j + \lambda_{\nu;ij}\bar{\nu}_iH_u\ell_j +
    \lambda_{d;ij}\bar{d}_iH_dq_j + \lambda_{e;ij}  \bar{e}_i  H_d\ell_j +
    \frac{1}{2}M_{R_i}\bar{\nu}_i\bar{\nu}_i \,.
\end{align}
The Yukawa matrices, $\lambda_{u,d,e,\nu}$, from this section onwards are defined in Weyl notation with doublets on the right, that is
\begin{align}
  \lambda_{u,d,e,\nu} = (Y^{u,d,e,\nu})^T \,.
\end{align}
We switched notation to doublets on the right because the renormalization group equations, in our program \texttt{maton}, are written with doublets on the right.

Without loss of generality, we can work in a basis where $\lambda_u, \lambda_d$, and $\lambda_e$ are diagonal.  Since we have chosen to work in the right-handed neutrino mass basis, we cannot simultaneously diagonalize $\lambda_\nu$.  Hence, the superpotential can be written as
\begin{align}
  \mathcal{W}
  = \sqrt{2}\alpha\Phi H_uH_d +
    \lambda_{u;ii}\bar{u}_iH_uq_i + \lambda_{\nu;ij}\bar{\nu}_iH_u\ell_j +
    \lambda_{d;ii}\bar{d}_iH_dq_i + \lambda_{e;ii}  \bar{e}_i  H_d\ell_i +
    \frac{1}{2}M_{R_i}\bar{\nu}_i\bar{\nu}_i \,.
\end{align}

From the superpotential, the $F$-term of $H_u$ is
\begin{align}
  |F_{H_u}|^2
    &= |\sqrt{2}\alpha\varphi h_d +
       \lambda_{u;ii}  \tilde{\bar{u}}_i  \tilde{q}_i       +
       \lambda_{\nu;ij}\tilde{\bar{\nu}}_i\tilde{\ell}_j|^2
  \nonumber
  \\&= 2\alpha^2|\varphi|^2|h_d|^2 +
       (\sqrt{2}\alpha\varphi h_d\lambda_{u;ii}^\dagger  \tilde{\bar{u}}_i^\dagger  \tilde{q}_i^\dagger    +
        \sqrt{2}\alpha\varphi h_d\lambda_{\nu;ij}^\dagger\tilde{\bar{\nu}}_i^\dagger\tilde{\ell}_j^\dagger +
        \text{h.c.}) + \dots \,,
\end{align}
where the ellipsis includes quartic sfermion terms.  The scalar component of the inflaton superfield is
\begin{align}
  \varphi = \frac{a+i\phi}{\sqrt{2}} \,,
\end{align}
where $\phi$ is the inflaton.  At the end of inflation, $a$ is stabilized at the origin and $\phi$ oscillates around $\phi=0$~\cite{Bryant:2016tzg}, we have
\begin{align}
  |F_{H_u}|^2
  = \alpha^2\phi^2|h_d|^2 +
    (i\lambda_{u;ii}^\dagger  \alpha\phi h_d\tilde{\bar{u}}_i^\dagger  \tilde{q}_i^\dagger    +
     i\lambda_{\nu;ij}^\dagger\alpha\phi h_d\tilde{\bar{\nu}}_i^\dagger\tilde{\ell}_j^\dagger +
     \text{h.c.}) + \dots \,.
  \label{eq:f_hu}
\end{align}
Similarly, the $F$-term of $H_d$ is
\begin{align}
  |F_{H_d}|^2
  = \alpha^2\phi^2|h_u|^2 +
    (i\lambda_{d;ii}^\dagger\alpha\phi h_u\tilde{\bar{d}}_i^\dagger\tilde{q}_i^\dagger    +
     i\lambda_{e;ii}^\dagger\alpha\phi h_u\tilde{\bar{e}}_i^\dagger\tilde{\ell}_i^\dagger +
     \text{h.c.}) + \dots \,.
  \label{eq:f_hd}
\end{align}
Another $F$-term that contributes to the production and the decay of the right-handed sneutrinos is
\begin{align}
  |F_{\bar{\nu}_i}|^2
  = \left|\sum_{j=1}^3
          \lambda_{\nu;ij}h_u \tilde{\ell}_j + M_{R_i}\tilde{\bar{\nu}}_i
    \right|^2
    = \left(\sum_{j=1}^3
          \lambda_{\nu;ij}^\dagger M_{R_i}h_u^\dagger \tilde{\ell}_j^\dagger\tilde{\bar{\nu}}_i +
          \text{h.c.}
    \right) + \dots \,.
\end{align}
where the ellipsis include quadratic and quartic scalar terms.  Hence, the Lagrangian includes
\begin{align}
  \begin{aligned}
    \mathcal{L}
    \supset \,\, &
    - \left( |F_{H_u}|^2 + |F_{H_d}|^2 +
             \sum_{i=1}^3|F_{\bar{\nu}_i}|^2 \right)
    \\ & -\left(
      \sqrt{2}\alpha\tilde{\phi}\tilde{h}_uh_d
    + \sqrt{2}\alpha\tilde{\phi}\tilde{h}_dh_u
    + \alpha\tilde{h}_u\tilde{h}_d\phi
    + \text{h.c.}\right)
    \\& -\bigg(
      \lambda_{u;ii}\bar{u}_ih_uq_i + \lambda_{\nu;ij}\bar{\nu}_ih_u\ell_j
    + \lambda_{d;ii}\bar{d}_ih_dq_i + \lambda_{d;ii}  \bar{e}_i  h_d\ell_i
    \\&\hspace{0.8cm}
    + \lambda_{u;ii}  \bar{u}_i  \tilde{h}_u\tilde{q}_i
    + \lambda_{\nu;ij}\bar{\nu}_i\tilde{h}_u\tilde{\ell}_j
    + \lambda_{d;ii}  \bar{d}_i  \tilde{h}_d\tilde{q}_i
    + \lambda_{d;ii}  \bar{e}_i  \tilde{h}_d\tilde{\ell}_i
    \\&\hspace{0.8cm}
    + \lambda_{u;ii}  \tilde{\bar{u}}_i  \tilde{h}_uq_i
    + \lambda_{\nu;ij}\tilde{\bar{\nu}}_i\tilde{h}_u\ell_j
    + \lambda_{d;ii}  \tilde{\bar{d}}_i  \tilde{h}_uq_i
    + \lambda_{d;ii}  \tilde{\bar{e}}_i  \tilde{h}_u\ell_i
    + \text{h.c.}\bigg)
    \,.
  \end{aligned}
  \label{eq:lag}
\end{align}

From the Lagrangian, we see that the Higgs masses are universal and time dependent:
\begin{align}
  m_h \equiv m_{h_u} = m_{h_d} = m_{\tilde{h}_u} = m_{\tilde{h}_d} = \alpha\langle\phi\rangle \,.
  \label{eq:mh}
\end{align}
Since the inflaton oscillation amplitude is of order the Planck scale, the Higgses can be heavier or lighter than the right-handed (s)neutrinos depending on the value of the inflaton vev.  Hence, the Higgses can decay to the right-handed (s)neutrinos and vice versa.

\subsection{Non-perturbative Decay of the Inflaton}
In this subsection, we provide a brief overview of instant preheating.  For detailed discussion please refer to Ref.~\cite{Kofman:1997yn,Felder:1998vq,Greene:2000ew}.

For a Lagrangian with the following term
\begin{align}
  \mathcal{L}
  \supset \frac{1}{2}\alpha^2\phi^2\chi^2 \,,
  \label{eq:broad_parametric}
\end{align}
where $\chi$ is a real scalar field, Kofman et.~al.~\cite{Kofman:1997yn} showed that when $\phi$ oscillates around $\phi=0$, $\phi$ creates $\chi$ states very efficiently at every zero-crossing.  The number density of $\chi$ created for a specific momentum $k$ is given by
\begin{align}
  n_k = \text{exp}\left(\frac{-\pi k^2}{\alpha|\dot{\phi}_0|}\right) \,,
  \label{eq:nk_created}
\end{align}
where $\dot{\phi}_0$ is the speed of $\phi$ at zero-crossing.  Hence, the number density of $\chi$ created at zero-crossing is
\begin{align}
  n_{\chi,0}
  = \int\frac{\mathrm{d}^3k}{(2\pi)^3}n_k
  = \frac{(\alpha|\dot{\phi}_0|)^{3/2}}{8\pi^3} \,,
  \label{eq:n_chi_created}
\end{align}
with a typical momentum of
\begin{align}
  k_\chi
  = \frac{1}{n_{\chi,0}}\int\frac{\mathrm{d}^3k}{(2\pi)^3}kn_k
  = \frac{2(\alpha|\dot{\phi}_0|)^{1/2}}{\pi} \,.
  \label{eq:ph}
\end{align}

In our model, the coupling between the two scalar Higgs doublets to the inflaton is of this form, therefore scalar Higgses are created efficiently at each zero-crossing with number density
\begin{align}
  n_{h_u,0} = n_{h_d,0} = 4n_{\chi,0} \,.
  \label{eq:nh0}
\end{align}
The factor of $4$ is because each Higgs doublet is complex and has four real degrees of freedom.

Similarly, for a Lagrangian that includes the following term,
\begin{align}
  \mathcal{L}
  \supset \alpha \ \phi \ \bar{\psi} \ {\psi} \,,
\end{align}
where $\psi$ is a fermion field, $\phi$ creates $\psi$ states very efficiently at every zero-crossing.  The number density of $\psi$ created is the same as in the bosonic case~\cite{Greene:2000ew}.  Hence, in our model, Higgsinos are also created efficiently at every zero-crossing with number density
\begin{align}
  n_{\tilde{h}_u,0}
  = n_{\tilde{h}_d,0}
  = 2n_\chi(0) \,.
  \label{eq:nht0}
\end{align}
The factor of $2$ is because the Higgsinos are doublets.

By conservation of energy, the inflaton speed is decreased by the following amount at every zero-crossing:
\begin{align}
  \Delta\dot{\phi}_0^2
  = 2\Delta\rho_\phi(0)
  = -2k_\chi[n_{h_u}(0)+n_{h_d}(0)+n_{\tilde{h}_u}(0)+n_{\tilde{h}_d}(0)]
  = -\frac{6\alpha^2|\dot{\phi}_0|^2}{\pi^4} \,.
  \label{eq:subtract_dphi}
\end{align}

It is important to note that eq.~(\ref{eq:nk_created}) is valid only if there are no background Higgses with momenta equal to the typical momentum in eq.~(\ref{eq:ph}).  Bose-Einstein effects from background Higgses with momentum equal to the typical momentum will further enhance the production rate~\cite{Kofman:1997yn}.  Since the thermalization rate of the Higgses, $\Gamma\sim n\sigma_\text{weak}\sim10^{20}\,\text{M}_\text{pl}$, is much bigger than the inflaton oscillation frequency $\sim10^{-6}\,\text{M}_\text{pl}$, the background Higgses, if they exist, are thermal.  Hence, out of the whole momentum spectrum, only Higgses with momenta close to the thermal temperature experience parametric enhancement.  To a good approximation, this parametric enhancement is ignored in this paper.

In addition, the non-perturbative instant preheating occurs while
\begin{align}
  q = \frac{\alpha^2\phi_\text{amp}^2}{4 m_\phi^2} \gg 1\,,
  \label{eq:q_cond}
\end{align}
and ends only when $q\sim1/3$~\cite{Kofman:1997yn}, where $\phi_\text{amp}$ is the inflaton oscillation amplitude and $m_\phi$ is the inflaton mass.

\subsection{Perturbative decay of the Inflaton}
In addition to the non-perturbative decay of the inflaton described in the preceding subsection, the inflaton can also decay perturbatively to the Higgses.  Although this effect is only significant long after broad parametric resonance ends, we include this effect at all times.  In all decay rates presented in this section and the next two, we use the approximation which neglects the mass of the decay products.  We guarantee energy conservation by including a Heaviside step function in all calculations.

The perturbative decay of the inflaton to the scalar Higgses is due to the $F$-terms of $H_u$ and $H_d$ in eq.~(\ref{eq:f_hu}) and (\ref{eq:f_hd}).  The decay rate of this process is given by
\begin{align}
  \Gamma_{\phi\to h}
  \equiv \Gamma_{\phi\to h_u} = \Gamma_{\phi\to h_d}
  = \frac{1}{16\pi}\frac{(\alpha m_h)^2}{m_\phi}
    \Theta(m_\phi-2m_h) \,.
  \label{eq:phi2h}
\end{align}
On the other hand, perturbative decay of the inflaton to the Higgsinos is due to the Yukawa-like terms in the Lagrangian in eq.~(\ref{eq:lag}).  The decay rate of this process is given by
\begin{align}
  \Gamma_{\phi\to\tilde{h}}
  \equiv \Gamma_{\phi\to\tilde{h}_u^0\tilde{h}_d^0} +
         \Gamma_{\phi\to\tilde{h}_u^+\tilde{h}_d^-}
  = 2\frac{1}{8\pi}\alpha^2m_\phi
    \Theta(m_\phi-2m_h) \,.
  \label{eq:phi2fh}
\end{align}

\subsection{Decay of the Higgses}
A very interesting phenomena of our model is that the scalar Higgses are massless when they are created non-perturbatively from the inflaton.  As the inflaton rolls up the potential, the Higgses obtain a mass proportional to the value of the inflaton vev shown in eq.~(\ref{eq:mh}).  Before the Higgses become heavier than the right-handed (s)neutrinos, they can only decay to radiation.  Eventually, the Higgses become massive enough and start decaying to the right-handed (s)neutrinos.  In addition, as we will see, the Higgs decay rates are proportional to their masses, that is the decay rates increase as the inflaton rolls up the potential.

\subsubsection{Up-type Higgses}
The possible decay channels of the up-type Higgses are
\begin{enumerate}
  \item Right-handed neutrinos: $h_u\to\bar{\nu}_i^\dagger \ell_j^\dagger$ with decay rate
  \begin{align}
    \begin{aligned}
      \Gamma_{h_u\to\bar{\nu}_i^\dagger}
        &= \sum_{j=1}^3
           \Gamma_{h_u        \to\bar{\nu}_i^\dagger\ell_j^\dagger} +
           \Gamma_{h_u^\dagger\to\bar{\nu}_i        \ell_j        }
         = \sum_{j=1}^3
           2\frac{1}{8\pi}|\lambda_{\nu;ij}|^2                     m_h
           \Theta(m_h-M_{R_i})
      \\&= 2\frac{1}{8\pi}(\lambda_{\nu}\lambda_{\nu}^\dagger)_{ii}m_h
           \Theta(m_h-M_{R_i}) \,.
      \label{eq:hu2n}
    \end{aligned}
  \end{align}
  The factor of $2$ is due to the charged and the neutral Higgses.

  \item Right-handed sneutrinos: $h_u\to\tilde{\bar{\nu}}_i^\dagger \tilde{\ell}_j^\dagger$ with decay rate
  \begin{align}
    \begin{aligned}
      \Gamma_{h_u\to\tilde{\bar{\nu}}_i^\dagger}
        &= \sum_{j=1}^3
           \Gamma_{h_u        \to\tilde{\bar{\nu}}_i^\dagger\tilde{\ell}_j^\dagger} +
           \Gamma_{h_u^\dagger\to\tilde{\bar{\nu}}_i        \tilde{\ell}_j        }
         = \sum_{j=1}^3
           2\frac{1}{16\pi}|\lambda_{\nu;ij}|^2                     \frac{M_{R_i}^2}{m_h}
           \Theta(m_h-M_{R_i})
      \\&= 2\frac{1}{16\pi}(\lambda_{\nu}\lambda_{\nu}^\dagger)_{ii}\frac{M_{R_i}^2}{m_h}
           \Theta(m_h-M_{R_i}) \,.
      \label{eq:hu2sn}
    \end{aligned}
  \end{align}

  \item Radiation: $h_u\to\tilde{\bar{d}}_i \tilde{q}_i, \tilde{\bar{e}}_i \tilde{\ell}_i, \bar{u}_i^\dagger q_i^\dagger$ with decay rate
  \begin{align}
    \Gamma_{h_u\to R}
    = \sum_{i=1}^3
      2\frac{1}{16\pi}(N_c|\lambda_{d;ii}|^2 + |\lambda_{e;ii}|^2)m_h +
      2\frac{1}{8\pi}  N_c|\lambda_{u;ii}|^2                      m_h   \,.
      \label{eq:hu2r}
  \end{align}

\end{enumerate}

\subsubsection{Down-type Higgses}
The possible decay channels of the down-type Higgses are
\begin{enumerate}
  \item Right-handed sneutrinos: $h_d\to\tilde{\bar{\nu}}_i \tilde{\ell}_j$ with decay rate
  \begin{align}
    \begin{aligned}
      \Gamma_{h_d\to\tilde{\bar{\nu}}_i}
        &= \sum_{j=1}^3
           \Gamma_{h_d        \to\tilde{\bar{\nu}}_i        \tilde{\ell}_j        } +
           \Gamma_{h_d^\dagger\to\tilde{\bar{\nu}}_i^\dagger\tilde{\ell}_j^\dagger}
         = \sum_{j=1}^3
           2\frac{1}{16\pi}|\lambda_{\nu;ij}|^2                     m_h
           \Theta(m_h-M_{R_i})
      \\&= 2\frac{1}{16\pi}(\lambda_{\nu}\lambda_{\nu}^\dagger)_{ii}m_h
           \Theta(m_h-M_{R_i}) \,.
      \label{eq:hd2sn}
    \end{aligned}
  \end{align}

  \item Radiation: $h_d\to\tilde{\bar{u}}_i \tilde{q}_i, \bar{d}_i^\dagger q_i^\dagger, \bar{e}_i^\dagger \ell_i^\dagger$ with decay rate
  \begin{align}
    \Gamma_{h_d\to R}
    = \sum_{i=1}^3
      2\frac{1}{16\pi} N_c|\lambda_{u;ii}|^2                      m_h +
      2\frac{1}{8 \pi}(N_c|\lambda_{d;ii}|^2 + |\lambda_{e;ii}|^2)m_h   \,.
      \label{eq:hd2r}
  \end{align}

\end{enumerate}

\subsubsection{Up-type Higgsinos}
The possible decay channels of the up-type Higgsinos are
\begin{enumerate}
  \item Right-handed neutrinos: $\tilde{h}_u\to\bar{\nu}_i^\dagger \tilde{\ell}_j^\dagger$ with decay rate
  \begin{align}
    \begin{aligned}
      \Gamma_{\tilde{h}_u\to\bar{\nu}_i^\dagger}
        &= \Gamma_{\tilde{h}_u        \to\bar{\nu}_i^\dagger\tilde{\ell}_j^\dagger} +
           \Gamma_{\tilde{h}_u^\dagger\to\bar{\nu}_i        \tilde{\ell}_j        }
         = \sum_{j=1}^3
           2\frac{1}{16\pi}|\lambda_{\nu;ij}|^2                     m_h
           \Theta(m_h-M_{R_i})
      \\&= 2\frac{1}{16\pi}(\lambda_{\nu}\lambda_{\nu}^\dagger)_{ii}m_h
           \Theta(m_h-M_{R_i}) \,.
      \label{eq:fhu2n}
    \end{aligned}
  \end{align}

  \item Right-handed sneutrinos: $\tilde{h}_u\to\tilde{\bar{\nu}}_i^\dagger \ell_j^\dagger$ with decay rate
  \begin{align}
    \begin{aligned}
      \Gamma_{\tilde{h}_u\to\tilde{\bar{\nu}}_i^\dagger}
        &= \sum_{j=1}^3
           \Gamma_{\tilde{h}_u        \to\tilde{\bar{\nu}}_i^\dagger\ell_j^\dagger} +
           \Gamma_{\tilde{h}_u^\dagger\to\tilde{\bar{\nu}}_i        \ell_j        }
         = \sum_{j=1}^3
           2\frac{1}{16\pi}|\lambda_{\nu;ij}|^2                     m_h
           \Theta(m_h-M_{R_i})
      \\&= 2\frac{1}{16\pi}(\lambda_{\nu}\lambda_{\nu}^\dagger)_{ii}m_h
           \Theta(m_h-M_{R_i}) \,.
      \label{eq:fhu2sn}
    \end{aligned}
  \end{align}
  \item Radiation: $\tilde{h}_u\to\bar{u}_i^\dagger \tilde{q}_i^\dagger, \tilde{\bar{u}}_i^\dagger q_i^\dagger$ with decay rate
  \begin{align}
    \Gamma_{\tilde{h}_u\to R}
    = \sum_{i=1}^3
      4\frac{1}{16\pi}N_c|\lambda_{u;ii}|^2m_h \,.
    \label{eq:fhu2r}
  \end{align}

\end{enumerate}

\subsubsection{Down-type Higgsinos}
The only decay channel of the down-type Higgsinos is to the radiation: $\tilde{h}_d\to\bar{d}_i^\dagger \tilde{q}_i^\dagger, \bar{e}_i^\dagger \tilde{\ell}_i^\dagger, \tilde{\bar{d}}_i^\dagger q_i^\dagger, \tilde{\bar{e}}_i^\dagger \ell_i^\dagger$ with decay rate
\begin{align}
  \Gamma_{\tilde{h}_d \to R}
  = \sum_{i=1}^3
    4\frac{1}{16\pi}(N_c|\lambda_{d;ii}|^2 + |\lambda_{e;ii}|^2)m_h \,.
  \label{eq:fhd2r}
\end{align}
This decay rate is multiplied by a factor of $4$ because the decay to $\bar{d}_i^\dagger\tilde{q}_i^\dagger$ and $\tilde{\bar{d}}_i^\dagger q_i^\dagger$ have the same coupling.  Similarly for the other two decay products.


\subsection{Decay of the Right-handed Neutrinos and Sneutrinos}
Right-handed (s)neutrinos can decay to Higgses(Higgsinos) and (s)leptons when they are heavier than the Higgses.

\subsubsection{Right-handed Neutrinos}
\label{ssec:neutrino}
Right-handed neutrinos can decay to
\begin{enumerate}
  \item Up-type Higgses: $\bar{\nu}_i\to\ell_j^\dagger h_u^\dagger$ with decay rate
  \begin{align}
    \begin{aligned}
      \Gamma_{\bar{\nu}_i\to h_u^\dagger}
        &= \sum_{j=1}^3
           \Gamma_{\bar{\nu}_i        \to\ell_j^\dagger h_u^\dagger} +
           \Gamma_{\bar{\nu}_i^\dagger\to\ell_j         h_u        }
         = \sum_{j=1}^3
           2\frac{1}{16\pi}|\lambda_{\nu;ij}|^2                   M_{R_i}
           \Theta(M_{R_i}-m_h)
      \\&= 2\frac{1}{16\pi}(\lambda_{\nu}\lambda_\nu^\dagger)_{ii}M_{R_i}
           \Theta(M_{R_i}-m_h) \,.
      \label{eq:n2hu}
    \end{aligned}
  \end{align}

  \item Up-type Higgsinos: $\bar{\nu}_i\to\tilde{\ell}_j^\dagger \tilde{h}_u^\dagger$ with decay rate
  \begin{align}
    \begin{aligned}
      \Gamma_{\bar{\nu}_i\to\tilde{h}_u^\dagger}
        &= \sum_{j=1}^3
           \Gamma_{\bar{\nu}_i        \to\tilde{\ell}_j^\dagger\tilde{h}_u^\dagger} +
           \Gamma_{\bar{\nu}_i^\dagger\to\tilde{\ell}_j        \tilde{h}_u        }
         = \sum_{j=1}^3
           2\frac{1}{16\pi}|\lambda_{\nu;ij}|^2                     M_{R_i}
           \Theta(M_{R_i}-m_h)
      \\&= 2\frac{1}{16\pi}(\lambda_{\nu}\lambda_{\nu}^\dagger)_{ii}M_{R_i}
           \Theta(M_{R_i}-m_h) \,.
      \label{eq:n2fhu}
    \end{aligned}
  \end{align}
\end{enumerate}

\subsubsection{Right-handed Sneutrinos}
\label{ssec:sneutrino}
Right-handed sneutrinos can decay to
\begin{enumerate}
  \item Up-type Higgses: $\tilde{\bar{\nu}}_i\to h_u^\dagger\tilde{\ell}_j^\dagger$ with decay rate
  \begin{align}
    \begin{aligned}
      \Gamma_{\tilde{\bar{\nu}}_i\to h_u^\dagger}
        &= \sum_{j=1}^3
           \Gamma_{\tilde{\bar{\nu}}_i        \to h_u^\dagger\tilde{\ell}_j^\dagger} +
           \Gamma_{\tilde{\bar{\nu}}_i^\dagger\to h_u        \tilde{\ell}_j        }
         = \sum_{j=1}^3
           2\frac{1}{16\pi}|\lambda_{\nu;ij}|^2                     M_{R_i}
           \Theta(M_{R_i}-m_h)
      \\&= 2\frac{1}{16\pi}(\lambda_{\nu}\lambda_{\nu}^\dagger)_{ii}M_{R_i}
           \Theta(M_{R_i}-m_h) \,.
      \label{eq:sn2hu}
    \end{aligned}
  \end{align}

  \item Down-type Higgses: $\tilde{\bar{\nu}}_i\to\tilde{\ell}_j^\dagger h_d$ with decay rate
  \begin{align}
    \begin{aligned}
    \Gamma_{\tilde{\bar{\nu}}_i\to h_d}
        &= \sum_{j=1}^3
           \Gamma_{\tilde{\bar{\nu}}_i        \to\tilde{\ell}_j^\dagger h_d        } +
           \Gamma_{\tilde{\bar{\nu}}_i^\dagger\to\tilde{\ell}_j         h_d^\dagger}
         = \sum_{j=1}^3
           2\frac{1}{16\pi}|\lambda_{\nu;ij}|^2                     \frac{m_h^2}{M_{R_i}}
           \Theta(M_{R_i}-m_h)
      \\&= 2\frac{1}{16\pi}(\lambda_{\nu}\lambda_{\nu}^\dagger)_{ii}\frac{m_h^2}{M_{R_i}}
           \Theta(M_{R_i}-m_h) \,.
      \label{eq:sn2hd}
    \end{aligned}
  \end{align}

  \item Up-type Higgsinos: $\tilde{\bar{\nu}}_i\to\ell_j^\dagger\tilde{h}_u^\dagger$ with decay rate
  \begin{align}
    \begin{aligned}
      \Gamma_{\tilde{\bar{\nu}}_i\to\tilde{h}_u^\dagger}
        &= \sum_{j=1}^3
           \Gamma_{\tilde{\bar{\nu}}_i        \to\ell_j^\dagger\tilde{h}_u^\dagger} +
           \Gamma_{\tilde{\bar{\nu}}_i^\dagger\to\ell_j        \tilde{h}_u        }
         = \sum_{j=1}^3
           2\frac{1}{16\pi}|\lambda_{\nu;ij}|^2                     M_{R_i}
           \Theta(M_{R_i}-m_h)
      \\&= 2\frac{1}{16\pi}(\lambda_{\nu}\lambda_{\nu}^\dagger)_{ii}M_{R_i}
           \Theta(M_{R_i}-m_h) \,.
    \end{aligned}
  \end{align}
\end{enumerate}

\subsection{Lepton Asymmetry}
A net lepton asymmetry can be produced when we consider the decay of the Higgses(Higgsinos) along with the subsequent decay of the right-handed (s)neutrinos.  Let the CP asymmetry of the Higgses(Higgsinos) decay be
\begin{align}
  \epsilon_{h_i}
  \equiv \frac{\Gamma_{h_u^\dagger\to\bar{\nu}_i        \ell        } -
               \Gamma_{h_u        \to\bar{\nu}_i^\dagger\ell^\dagger}   }
              {\Gamma_{h_u^\dagger\to\bar{\nu}_i        \ell        } +
               \Gamma_{h_u        \to\bar{\nu}_i^\dagger\ell^\dagger}   } \,,
  \label{eq:epsi}
\end{align}
and that of the right-handed (s)neutrinos decay be
\begin{align}
  \epsilon_{\bar{\nu}_i}
  \equiv \frac{\Gamma_{\bar{\nu}_i^\dagger\to\ell         h_u        } -
               \Gamma_{\bar{\nu}_i        \to\ell^\dagger h_u^\dagger}   }
              {\Gamma_{\bar{\nu}_i^\dagger\to\ell         h_u        } +
               \Gamma_{\bar{\nu}_i        \to\ell^\dagger h_u^\dagger}   } \,,
\end{align}
where the family indices of the leptons are summed.  Then, for example, when an up-type Higgs decay, we have
\begin{align}
  h_u \to
  \begin{cases}
      \dfrac{1+\epsilon_{h_i}}{2}\,\bar{\nu}_i      \ell         \to&
      \begin{cases}
          \dfrac{(1+\epsilon_{h_i})(1+\epsilon_{\bar{\nu}_i})}{4}\,h_u        \ell        \ell
        \\\dfrac{(1+\epsilon_{h_i})(1-\epsilon_{\bar{\nu}_i})}{4}\,h_u^\dagger\ell^\dagger\ell
      \end{cases}
    \\\dfrac{1-\epsilon_{h_i}}{2}\bar{\nu}_i^\dagger\ell^\dagger \to&
      \begin{cases}
          \dfrac{(1-\epsilon_{h_i})(1+\epsilon_{\bar{\nu}_i})}{4}h_u        \ell        \ell^\dagger
        \\\dfrac{(1-\epsilon_{h_i})(1-\epsilon_{\bar{\nu}_i})}{4}h_u^\dagger\ell^\dagger\ell^\dagger
      \end{cases}
  \end{cases} \,,
\end{align}
where the $\epsilon$ factors are the branching ratios.  We see that only half of the decay channels have a net lepton asymmetry.  Hence, the final lepton asymmetry is
\begin{align}
  n_{L}
  \equiv n_{\ell}-n_{\bar{\ell}}
  = 2\frac{(1+\epsilon_{h_i})(1+\epsilon_{\bar{\nu}_i})}{4}n_{h_u} -
    2\frac{(1-\epsilon_{h_i})(1-\epsilon_{\bar{\nu}_i})}{4}n_{h_u}
  = \epsilon_{h_i}        n_{h_u}         +
    \epsilon_{\bar{\nu}_i}n_{\bar{\nu}_i}   \,.  \label{eqn:nL}
\end{align}
In the last equality, we used $n_{\bar{\nu}_i}=n_{h_u}$, which is true in this process because each up-type Higgs creates a right-handed neutrino.  There is a factor of $2$ multiplying the branching ratio of the lepton asymmetric final states because these states have either two leptons or two anti-leptons\footnote{In our evaluation of the net lepton asymmetry we follow the analysis of Ahn and Kolb~\cite{Ahn:2005bg}.  We note here that, with regards to their formula equivalent to eq.~(\ref{eqn:nL}), they do not take into account that the final states have either two leptons or two anti-leptons, therefore our lepton asymmetry is a factor of $2$ bigger than theirs.}.

The CP asymmetry of the right-handed neutrinos is given by~\cite{Covi:1996wh,Buchmuller:1997yu,Giudice:2003jh,Buchmuller:2005eh,Davidson:2008bu}
\begin{align}
  \epsilon_{\bar{\nu}_i}
  = - \frac{1}{8\pi}\sum_{j \neq i} \frac{\text{Im}\{[(\lambda_\nu \lambda_\nu^\dagger)_{ji}]^2\}}{(\lambda_\nu \lambda_\nu^\dagger)_{ii}}g\left(\frac{M_{R_j}}{M_{R_i}}\right) \,,
  \label{eq:eps}
\end{align}
where
\begin{align}
  g(x) = - \sqrt{x}\left(\frac{2}{x-1}+\ln\frac{1+x}{x}\right) \,.
\end{align}
Note, our result agrees with the sign of the results in Refs.~\cite{Covi:1996wh,Giudice:2003jh,Buchmuller:2005eh}, but disagree with the sign in Ref.~\cite{Davidson:2008bu}.  In order to compare equations, one needs to use the dictionary relating the different definitions of Yukawa matrices given in App.~\ref{app:yukawas}.  In addition, we do not use the limiting form of this equation for $x\gg1$ in our calculation because we also consider the case when $x\ll1$.

Eq.~(\ref{eq:eps}) is calculated when the $j^\text{th}$ right-handed (s)neutrinos are present in the loop.  However, when the second lightest right-handed (s)neutrinos decay, the heaviest right-handed (s)neutrino is already integrated out from the model to give us the Weinberg operator (for $n = 3$)
\begin{align}
  \eta^{(n)}_{i j} = \sum_{k = \{ n \}} {\lambda_\nu^T}_{i k}\frac{1} M_{R_k} {\lambda_\nu}_{k j} \,.
\end{align}
Hence, instead of using eq.~(\ref{eq:eps}), the CP asymmetry parameter should be calculated using the Weinberg operator~\cite{Buchmuller:2000nd}
\begin{align}
  \epsilon_{\bar{\nu}_i}
  \sim \frac{3}{8\pi}\frac{\text{Im}[(\lambda_\nu^*\eta^{(n)}\lambda_\nu^\dagger)_{ii}]}{(\lambda_\nu \lambda_\nu^\dagger)_{ii}}M_{R_i} \,.
\end{align}
$\epsilon_{\bar{\nu}_i}$ in Ref.~\cite{Buchmuller:2000nd} has a factor of $16\pi$ instead of $8\pi$ because the asymmetry parameter is calculated for the SM.

To summarize, the CP asymmetry parameters for the heaviest right-handed (s)neutrinos, $\bar{\nu}_3$, to the lightest right-handed (s)neutrinos, $\bar{\nu}_1$, are given by
\begin{align}
  \begin{aligned}
    \epsilon_{\bar{\nu}_3} = \epsilon_{h_3}
    =&\, - \frac{1}{8\pi}\sum_{j=1,2}\frac{\text{Im}\{[(\lambda_\nu \lambda_\nu^\dagger)_{j3}]^2\}}{(\lambda_\nu \lambda_\nu^\dagger)_{33}}g\left(\frac{M_{R_j}}{M_{R_3}}\right)
    \\\epsilon_{\bar{\nu}_2} = \epsilon_{h_2}
    =&\, - \frac{1}{8\pi}\frac{\text{Im}\{[(\lambda_\nu \lambda_\nu^\dagger)_{12}]^2\}}{(\lambda_\nu \lambda_\nu^\dagger)_{22}}g\left(\frac{M_{R_1}}{M_{R_2}}\right)
    + \frac{3}{8\pi}\frac{\text{Im}[(\lambda_\nu^*\eta^{(3)}\lambda_\nu^\dagger)_{22}]}{(\lambda_\nu \lambda_\nu^\dagger)_{22}}M_{R_2}
    \\\epsilon_{\bar{\nu}_1} = \epsilon_{h_1}
    =&\, \frac{3}{8\pi}\frac{\text{Im}[(\lambda_\nu^*\eta^{(2,3)}\lambda_\nu^\dagger)_{11}]}{(\lambda_\nu \lambda_\nu^\dagger)_{11}}M_{R_1} \,,
  \end{aligned}
\end{align}
where we have made the assumption that the decay products are massless.  With this assumption, the CP asymmetries due to the Higgses decay and the right-handed (s)neutrinos decay are equal~\cite{Ahn:2005bg}.  This assumption is made throughout the paper.

An interesting feature of our model is that the phases in the right-handed neutrino Yukawa matrix, $\lambda_\nu = {Y^\nu}^T$, are fixed by fitting to the low energy data.  With these phases, the decay of the heaviest right-handed (s)neutrinos produce more anti-leptons than leptons, while the decay of the two lighter right-handed (s)neutrinos produce more leptons than anti-leptons.  Hence, unlike most models in the literature, we cannot integrate out any right-handed (s)neutrinos.  Most of our baryon asymmetry is created from the heaviest right-handed (s)neutrinos, while the two lighter right-handed (s)neutrinos wash out a portion of the asymmetry.

\subsection{Decay of the Waterfall Fields}
In addition to the inflaton, there are waterfall fields, $\sigma$, after the inflation ends.  The relevant superpotential is given by eq.~(\ref{eq:Wnu}) with $\lambda_1 \equiv \lambda_2$.  The $F$-term of $Q^c$ is
\begin{align}
  \begin{aligned}
    |F_{Q_i^c}|^2
    &=         \left|\frac{\lambda_i^2}{M_i}\left(\frac{\sigma+i\tau+\sqrt{2}v_\text{PS}}{2}\right)^2\tilde{\bar{\nu}}_i\right|^2
    \\&\supset \frac{\lambda_i^4}{M_i^2}v_\text{PS}^2\tilde{\bar{\nu}}_i\tilde{\bar{\nu}}_i^\dagger
               \left(\frac{3}{4}\sigma^2 +
                     \frac{1}{\sqrt{2}}v_\text{PS}\sigma +
                     \frac{1}{4}v_\text{PS}^2\right)
    \\&=   \frac{3}{2}h_i^2\sigma^2\tilde{\bar{\nu}}_i\tilde{\bar{\nu}}_i^\dagger
         + \sqrt{2}h_i^2v_\text{PS}\sigma\tilde{\bar{\nu}}_i\tilde{\bar{\nu}}_i^\dagger
         + M_{R_i}^2\tilde{\bar{\nu}}_i\tilde{\bar{\nu}}_i^\dagger \,,
  \end{aligned}
\end{align}
where $h_i$ and $M_{R_i}$ are given in eq.~(\ref{eq:MRi_hi}).  The first term in this $F$-term is similar to the broad parametric resonance term in eq.~(\ref{eq:broad_parametric}).  Hence, one would expect parametric resonance to occur in the waterfall field.  However, the corresponding broad parametric resonance parameter is too small for broad parametric resonance to occur
\begin{align}
  q = \frac{9h_i^4\sigma_\text{amp}}{4m_\sigma^2} \ll 1 \,.
\end{align}

On the other hand, the second term in the $F$-term above allows for the decay of waterfall field to right-handed sneutrinos, while the Yukawa-like terms from the superpotential in eq.~(\ref{eq:sdecay}) allows for the decay of waterfall fields to right-handed neutrinos.  Hence, the waterfall field perturbative decay rates are
\begin{align}
  \Gamma_{\sigma\to\bar{\nu}_i\bar{\nu}_i}
  =&\, \frac{1}{64\pi}h_i^2m_\sigma
  \\\Gamma_{\sigma\to\tilde{\bar{\nu}}_i\tilde{\bar{\nu}}_i}
  =&\, \frac{1}{16 \pi}\frac{h_i^4v_\text{PS}^2}{m_\sigma} \,.
\end{align}
Moreover, since $h_i=\sqrt{2}M^i_R/v_{PS}$,  the waterfall field decays predominantly to the heaviest right-handed (s)neutrinos.

\section{Evolution Equations}
\label{sec:evolution}
To analyze the evolution of all particles after inflation ends, we follow the approach used by Ahn et.~al.~\cite{Ahn:2005bg}.  As a first approximation, we do not consider the momentum of the particles.

\subsection{Equation of Motion of the Inflaton}
The equation of motion of the inflaton field is given by~\cite{Kofman:1997yn}
\begin{align}
  \begin{aligned}
    \ddot{\phi}+3H\dot{\phi}+m_\phi^2\phi+\alpha^2\langle h_u^2+h_d^2+\tilde{h}_u^2+\tilde{h}_d^2\rangle\phi
   =&\, - 2\Gamma_{\phi\to h}       \dot{\phi}
        -  \Gamma_{\phi\to\tilde{h}}\dot{\phi} \,,
  \end{aligned}
\end{align}
where by the Hartree approximation, defined in Ref.~\cite{Kofman:1997yn}, the vev of the Higgses is of the form
\begin{align}
  \langle h^2\rangle = \frac{n_h}{m_h} = \frac{n_h}{\alpha|\phi|} \,.
\end{align}
The factor of $2$ multiplying the decay rate is because the inflaton can decay to both the up-type and the down-type Higgses.  Hence, the inflaton equation of motion can be written as
\begin{align}
  \begin{aligned}
    \ddot{\phi}+3H\dot{\phi}+m_\phi^2\phi+\alpha(n_{h_u}+n_{h_d}+n_{\tilde{h}_u}+n_{\tilde{h}_d})\text{sign}(\phi)
   =&\, - 2\Gamma_{\phi\to h}       \dot{\phi}
        -  \Gamma_{\phi\to\tilde{h}}\dot{\phi} \,,
  \end{aligned}
  \label{eq:eom_phi_early}
\end{align}
where $H$ is the Hubble parameter in units of the reduced Planck mass,
\begin{align}
  H^2 = \frac{1}{3}\left[\rho_\phi+m_h(n_{h_u}+n_{h_d}+n_{\tilde{h}_u}+n_{\tilde{h}_d})+\sum_iM_{R_i}(n_{\bar{\nu}_i}+n_{\tilde{\bar{\nu}}_i})+\rho_R\right] \,.
  \label{eq:h}
\end{align}

\subsection{Evolution Equations for Number Density of Higgses}
To derive the evolution equation for the Higgses, we start by considering just the interaction between the inflaton and the Higgses.  The inflaton energy density is defined as
\begin{align}
  \rho_\phi = \frac{1}{2}\dot{\phi}^2 + \frac{1}{2}m_{\phi}^2\phi^2 \,.
\end{align}
With this definition, the rate of change of the inflaton energy density is
\begin{align}
  \dot{\rho}_\phi = \dot{\phi}(\ddot{\phi}+m_{\phi}^2\phi) \,.
\end{align}
By multiplying the equation of motion with $\dot{\phi}$, we have
\begin{align}
  \begin{aligned}
    \dot{\rho}_\phi+3H\dot{\phi}^2+\dot{m}_h(n_{h_u}+n_{h_d}+n_{\tilde{h}_u}+n_{\tilde{h}_d})
   =&\, - 2\Gamma_{\phi\to h}       \dot{\phi}^2
        -  \Gamma_{\phi\to\tilde{h}}\dot{\phi}^2
  \end{aligned}
\end{align}
To conserve energy between the inflaton and the Higgses, we have
\begin{align}
  \begin{aligned}
        \dot{\rho}_{h_u}+3H\rho_{h_u}
      - \dot{m}_hn_{h_u}
      - \Gamma_{\phi\to h}\dot{\phi}^2
     &= 0 \,.
  \end{aligned}
\end{align}
With $\rho_h = E_hn_h$, we have
\begin{align}
  \begin{aligned}
        E_h\dot{n}_{h_u}+\dot{E}_hn_{h_u}+3HE_hn_{h_u}
      - \dot{m}_hn_{h_u}
      - \Gamma_{\phi\to h}\dot{\phi}^2
     &= 0
    \\  \dot{n}_{h_u}+3Hn_{h_u}
      + \frac{\dot{E}_h-\dot{m}_h}{E_h}n_{h_u}
      - \Gamma_{\phi\to h}\frac{\dot{\phi}^2}{E_h}
     &= 0
  \end{aligned}
\end{align}
Since we are ignoring the momentum of the Higgses, $\dot{E}_h=\dot{m}_h$.  In addition, conservation of energy requires $m_\phi=2E_h$.  So, the evolution equation of the Higgses that conserves energy with that of the inflaton is given by
\begin{align}
  \begin{aligned}
        \dot{n}_{h_u}+3Hn_{h_u}
      - 2\Gamma_{\phi\to h}\frac{\dot{\phi}^2}{m_\phi}
     &= 0
  \end{aligned} \,.
\end{align}
The evolution equations for the three other Higgses are very similar and we will omit the derivation.

Including the decay and the creation of the Higgses, the full evolution equations for the Higgses are given by
\begin{align}
  &\begin{aligned}
    \dot{n}_{h_u} + 3Hn_{h_u}
    = &\, \sum_{i=1}^3 \bigg[ - \gamma^{-1}_h            \Gamma_{h_u\to\bar{\nu}_i^\dagger        }(n_{h_u}-n_h^\text{eq})
                              - \gamma^{-1}_h            \Gamma_{h_u\to\tilde{\bar{\nu}}_i^\dagger}(n_{h_u}-n_h^\text{eq})
    \\&\hspace{1.04cm}        + \gamma^{-1}_{\bar{\nu}_i}\Gamma_{\bar{\nu}_i\to h_u^\dagger        }n_{\bar{\nu}_i}
                              + \gamma^{-1}_{\bar{\nu}_i}\Gamma_{\tilde{\bar{\nu}}_i\to h_u^\dagger}n_{\tilde{\bar{\nu}}_i} \bigg]
    \\&\, - \gamma^{-1}_h\Gamma_{h_u\to R}(n_{h_u}-n_h^\text{eq})
          + 2\Gamma_{\phi\to h}\frac{\dot{\phi}^2}{m_\phi}
  \end{aligned}
  \label{eq:ee_hu}
  \\&
  \begin{aligned}
    \dot{n}_{h_d} + 3Hn_{h_d}
    = &\, \sum_{i=1}^3 \bigg[ - \gamma^{-1}_h            \Gamma_{h_d\to\tilde{\bar{\nu}}_i}(n_{h_d}-n_h^\text{eq})
                              + \gamma^{-1}_{\bar{\nu}_i}\Gamma_{\tilde{\bar{\nu}}_i\to h_d}n_{\tilde{\bar{\nu}}_i} \bigg]
    \\&\, - \gamma^{-1}_h\Gamma_{h_d\to R}(n_{h_d}-n_h^\text{eq})
          + 2\Gamma_{\phi\to h}\frac{\dot{\phi}^2}{m_\phi}
  \end{aligned}
  \label{eq:ee_hd}
  \\&
  \begin{aligned}
    \dot{n}_{\tilde{h}_u} + 3Hn_{\tilde{h}_u}
    = &\, \sum_{i=1}^3 \bigg[ - \gamma^{-1}_h            \Gamma_{\tilde{h}_u\to\bar{\nu}_i^\dagger        }(n_{\tilde{h}_u}-n_h^\text{eq})
                              - \gamma^{-1}_h            \Gamma_{\tilde{h}_u\to\tilde{\bar{\nu}}_i^\dagger}(n_{\tilde{h}_u}-n_h^\text{eq})
    \\&\hspace{1.04cm}        + \gamma^{-1}_{\bar{\nu}_i}\Gamma_{\bar{\nu}_i\to\tilde{h}_u^\dagger        }n_{\bar{\nu}_i}
                              + \gamma^{-1}_{\bar{\nu}_i}\Gamma_{\tilde{\bar{\nu}}_i\to\tilde{h}_u^\dagger}n_{\tilde{\bar{\nu}}_i}          \bigg]
    \\&\, - \gamma^{-1}_h\Gamma_{\tilde{h}_u\to R}(n_{\tilde{h}_u}-n_h^\text{eq})
          + 2\Gamma_{\phi\to\tilde{h}}\frac{\dot{\phi}^2}{m_{\phi}}
  \end{aligned}
  \label{eq:ee_hut}
  \\&
  \begin{aligned}
    \dot{n}_{\tilde{h}_d} + 3Hn_{\tilde{h}_d}
    = &\, -  \gamma^{-1}_h\Gamma_{\tilde{h}_d\to R}(n_{\tilde{h}_d}-n_h^\text{eq})
          + 2\Gamma_{\phi\to\tilde{h}}\frac{\dot{\phi}^2}{m_{\phi}} \,.
  \end{aligned}
  \label{eq:ee_hdt}
\end{align}
Since these evolution equations do not take momentum into account, we need to explicitly enforce detailed balance.  We thus replace $n_h$ by $(n_h-n_h^{\text{eq}})$, where $n_h^\text{eq}$ is the thermal equilibrium number density of the Higgses, because the Higgs number densities should always approach the equilibrium number.  The thermal equilibrium number density of the Higgses are given by
\begin{align}
  n_h^\text{eq}
  = \frac{g}{2\pi^3}\int_{m_h}^\infty\mathrm{d}E\,\,\,E\sqrt{E^2-m_h^2}e^{-E/T}
  = \frac{g}{2\pi^3}T^3\frac{m_h^2}{T^2}K_2\left(\frac{m_h}{T}\right) \,,
\end{align}
where $K_2$ is the modified Bessel function of the second kind, the internal degrees of freedom (the $g$-factor) of the Higgses is $g=4$ and we have used Maxwell-Boltzmann statistics.  Since the masses and $g$-factor are the same among the Higgses, they have the same thermal equilibrium number density.  On the other hand, the (s)neutrinos never have sufficient time to equilibrate.

Both Higgses and right-handed (s)neutrinos are in kinetic equilibrium with the thermal bath when the thermal temperature is larger than their corresponding masses.  Thus, we have to take into account that Higgses and right-handed (s)neutrinos are not decaying at rest by multiplying their decay rates by a dilation factor, $\gamma$, where $\gamma_h=\sqrt{m_h^2+T^2}/m_h$ and $\gamma_{\bar{\nu}_i}=\sqrt{m_{\bar{\nu}_i}^2+T^2}/m_{\bar{\nu}_i}$.\footnote{We would like to thank Gary Steigman for pointing this out.}

\subsection{Evolution Equations for the Number density of Right-handed Neutrinos and Sneutrinos}
The evolution equations for the right-handed (s)neutrinos are
\begin{align}
  &
  \begin{aligned}
    \dot{n}_{\bar{\nu}_i} + 3Hn_{\bar{\nu}_i}
     =&\, - \gamma^{-1}_{\bar{\nu}_i}\Gamma_{\bar{\nu}_i\to h_u^\dagger       }n_{\bar{\nu}_i}
          - \gamma^{-1}_{\bar{\nu}_i}\Gamma_{\bar{\nu}_i\to\tilde{h}_u^\dagger}n_{\bar{\nu}_i}
    \\&\, + \gamma^{-1}_h            \Gamma_{h_u        \to\bar{\nu}_i^\dagger}(n_{h_u}-n_h^\text{eq})
          + \gamma^{-1}_h            \Gamma_{\tilde{h}_u\to\bar{\nu}_i^\dagger}(n_{\tilde{h}_u}-n_h^\text{eq})
    \\&\, + 2\Gamma_{\sigma\to\bar{\nu}_i}\frac{\rho_\sigma}{m_{\sigma}}
  \end{aligned}
  \label{eq:ee_nu}
  \\&
  \begin{aligned}
    \dot{n}_{\tilde{\bar{\nu}}_i} + 3Hn_{\tilde{\bar{\nu}}_i}
    =& \, - \gamma^{-1}_{\bar{\nu}_i}\Gamma_{\tilde{\bar{\nu}}_i\to h_u^\dagger       }n_{\tilde{\bar{\nu}}_i}
          - \gamma^{-1}_{\bar{\nu}_i}\Gamma_{\tilde{\bar{\nu}}_i\to h_d               }n_{\tilde{\bar{\nu}}_i}
          - \gamma^{-1}_{\bar{\nu}_i}\Gamma_{\tilde{\bar{\nu}}_i\to\tilde{h}_u^\dagger}n_{\tilde{\bar{\nu}}_i}
    \\&\, + \gamma^{-1}_h            \Gamma_{h_u        \to\tilde{\bar{\nu}}_i^\dagger}(n_{h_u}-n_h^\text{eq})
          + \gamma^{-1}_h            \Gamma_{h_d        \to\tilde{\bar{\nu}}_i        }(n_{h_d}-n_h^\text{eq})
    \\&\, + \gamma^{-1}_h            \Gamma_{\tilde{h}_u\to\tilde{\bar{\nu}}_i^\dagger}(n_{\tilde{h}_u}-n_h^\text{eq})
    \\&\, + 2\Gamma_{\sigma\to\tilde{\bar{\nu}}_i}\frac{\rho_\sigma}{m_{\sigma}} \,.
  \end{aligned}
  \label{eq:ee_nut}
\end{align}

\subsection{Evolution Equation for Energy Density of Waterfall Fields}
The evolution equation for the energy density of waterfall fields is
\begin{align}
  \dot{\rho}_\sigma + 3H\rho_\sigma + \Gamma_{\sigma\to\bar{\nu}_i}\rho_\sigma + \Gamma_{\sigma\to\tilde{\bar{\nu}}_i}\rho_\sigma = 0 \,.
  \label{eq:ee_sigma}
\end{align}

\subsection{Evolution Equation for the Energy Density of Radiation}
The evolution equation for the energy density of radiation is\footnote{Recall, $M_{R_i}$ is the mass of the heavy right-handed (s)neutrinos.  Until the inflaton field settles at its minimum $\langle\phi\rangle=0$, the Higgs are on average very heavy.}
\begin{align}
  \begin{aligned}
    \dot{\rho}_R + 4H\rho_R
    = \sum_{i=1}^3 &\,
          + \frac{1}{2}\gamma^{-1}_h            \Gamma_{h_u\to\bar{\nu}_i^\dagger        }m_h(n_{h_u}-n_h^\text{eq})
          + \frac{1}{2}\gamma^{-1}_h            \Gamma_{h_u\to\tilde{\bar{\nu}}_i^\dagger}m_h(n_{h_u}-n_h^\text{eq})
    \\&\, +            \gamma^{-1}_h            \Gamma_{h_u\to R                         }m_h(n_{h_u}-n_h^\text{eq})
    \\&\, + \frac{1}{2}\gamma^{-1}_h            \Gamma_{h_d\to\tilde{\bar{\nu}}_i        }m_h(n_{h_d}-n_h^\text{eq})
          +            \gamma^{-1}_h            \Gamma_{h_d\to R                         }m_h(n_{h_d}-n_h^\text{eq})
    \\&\, + \frac{1}{2}\gamma^{-1}_h            \Gamma_{\tilde{h}_u\to\bar{\nu}_i^\dagger        }m_h(n_{\tilde{h}_u}-n_h^\text{eq})
          + \frac{1}{2}\gamma^{-1}_h            \Gamma_{\tilde{h}_u\to\tilde{\bar{\nu}}_i^\dagger}m_h(n_{\tilde{h}_u}-n_h^\text{eq})
    \\&\, +            \gamma^{-1}_h            \Gamma_{\tilde{h}_u\to R                         }m_h(n_{\tilde{h}_u}-n_h^\text{eq})
    \\&\, +            \gamma^{-1}_h            \Gamma_{\tilde{h}_d\to R                         }m_h(n_{\tilde{h}_d}-n_h^\text{eq})
    \\&\, + \frac{1}{2}\gamma^{-1}_{\bar{\nu}_i}\Gamma_{\bar{\nu}_i        \to h_u^\dagger       }M_{R_i}n_{\bar{\nu}_i}
          + \frac{1}{2}\gamma^{-1}_{\bar{\nu}_i}\Gamma_{\bar{\nu}_i        \to\tilde{h}_u^\dagger}M_{R_i}n_{\bar{\nu}_i}
    \\&\, + \frac{1}{2}\gamma^{-1}_{\bar{\nu}_i}\Gamma_{\tilde{\bar{\nu}}_i\to h_u^\dagger       }M_{R_i}n_{\tilde{\bar{\nu}}_i}
          + \frac{1}{2}\gamma^{-1}_{\bar{\nu}_i}\Gamma_{\tilde{\bar{\nu}}_i\to h_d               }M_{R_i}n_{\tilde{\bar{\nu}}_i}
          + \frac{1}{2}\gamma^{-1}_{\bar{\nu}_i}\Gamma_{\tilde{\bar{\nu}}_i\to\tilde{h}_u^\dagger}M_{R_i}n_{\tilde{\bar{\nu}}_i} \,.
  \end{aligned}
  \label{eq:ee_r}
\end{align}

\subsection{Evolution Equation for the Number Density of Lepton Asymmetry}
The number density of lepton asymmetry is defined by
\begin{align}
  n_L = n_{\ell}-n_{\bar{\ell}} \,,
\end{align}
and its evolution equation is
\begin{align}
  \begin{aligned}
    \dot{n}_L + 3Hn_L
    = \sum_{i=1}^3 &\,
          + \epsilon_{h_i}        \gamma^{-1}_h            \Gamma_{h_u\to\bar{\nu}_i^\dagger        }(n_{h_u}-n_h^\text{eq})
          + \epsilon_{h_i}        \gamma^{-1}_h            \Gamma_{h_u\to\tilde{\bar{\nu}}_i^\dagger}(n_{h_u}-n_h^\text{eq})
    \\&\, + \epsilon_{h_i}        \gamma^{-1}_h            \Gamma_{h_d\to\tilde{\bar{\nu}}_i        }(n_{h_d}-n_h^\text{eq})
    \\&\, + \epsilon_{h_i}        \gamma^{-1}_h            \Gamma_{\tilde{h}_u\to\bar{\nu}_i^\dagger        }(n_{\tilde{h}_u}-n_h^\text{eq})
          + \epsilon_{h_i}        \gamma^{-1}_h            \Gamma_{\tilde{h}_u\to\tilde{\bar{\nu}}_i^\dagger}(n_{\tilde{h}_u}-n_h^\text{eq})
    \\&\, + \epsilon_{\bar{\nu}_i}\gamma^{-1}_{\bar{\nu}_i}\Gamma_{\bar{\nu}_i        \to h_u^\dagger       }n_{\bar{\nu}_i}
          + \epsilon_{\bar{\nu}_i}\gamma^{-1}_{\bar{\nu}_i}\Gamma_{\bar{\nu}_i        \to\tilde{h}_u^\dagger}n_{\bar{\nu}_i}
    \\&\, + \epsilon_{\bar{\nu}_i}\gamma^{-1}_{\bar{\nu}_i}\Gamma_{\tilde{\bar{\nu}}_i\to h_u^\dagger       }n_{\tilde{\bar{\nu}}_i}
          + \epsilon_{\bar{\nu}_i}\gamma^{-1}_{\bar{\nu}_i}\Gamma_{\tilde{\bar{\nu}}_i\to h_d               }n_{\tilde{\bar{\nu}}_i}
          + \epsilon_{\bar{\nu}_i}\gamma^{-1}_{\bar{\nu}_i}\Gamma_{\tilde{\bar{\nu}}_i\to\tilde{h}_u^\dagger}n_{\tilde{\bar{\nu}}_i} \,,
  \end{aligned}
  \label{eq:ee_nl}
\end{align}
where $\epsilon_{h_i}$ is defined in eq.~(\ref{eq:epsi}).

\section{Procedure}
\label{sec:procedure}
We start the analysis of the reheating and the baryogenesis at the first zero-crossing after the last 60 e-foldings of inflation. The details of the inflationary epoch can be found in Ref.~\cite{Bryant:2016tzg}.

Unless otherwise specified, the parameters used to produce the results in the rest of the paper are as follows:
\begin{equation}
  \begin{gathered}
    \phi(0) = 0 \,,
    \hspace{1cm}
    \dot{\phi}(0) = -2.72\times10^{-6}\,\text{M}_{\text{pl}}^2 \,,
    \\
    m_{\phi} = 5.80\times10^{-6}\,\text{M}_{\text{pl}} \,,
    \hspace{1cm}
    m_{\sigma} = 9.80\times10^{-3}\,\text{M}_{\text{pl}} \,,
    \\
    \rho_\sigma(0) = 5.18\times10^{-15}\,\text{M}_{\text{pl}}^4 \,,
    \hspace{1cm}
    h_{3,2,1}^2 = 2\frac{M_{R_i}^2}{v_\text{PS}^2} \,,
    \\\sum_{i=1}^3|\lambda_{u;ii}|^2 = 0.4099 \,,
    \hspace{0.6cm}
    \sum_{i=1}^3|\lambda_{d;ii}|^2 = 0.4041 \,,
    \hspace{0.6cm}
    \sum_{i=1}^3|\lambda_{e;ii}|^2 = 0.3245 \,,
    \hspace{0.6cm}
    \\M_{R_{3,2,1}} = \{1.136\times10^{-5},2.337\times10^{-7},3.685\times10^{-9}\}\,\text{M}_{\text{pl}} \,,
    \\
    (\lambda_{\nu}\lambda_{\nu}^\dagger)_{33,22,11} = \{3.316\times10^{-1},1.255\times10^{-2},1.736\times10^{-4}\} \,,
    \\
    \epsilon_{3,2,1} = \{-4.429\times10^{-5},6.038\times10^{-4},6.044\times10^{-7}\} \,.
  \end{gathered}
  \label{eq:parameters}
\end{equation}
The inflation parameters are obtained by fitting to inflation observables~\cite{Bryant:2016tzg} while the matter sector parameters are obtained by fitting to the low-energy observables.  A benchmark point of low-energy fits is given in App.~\ref{app:low_energy}.

\subsection{Non-perturbative regime}
\label{sec:np_regime}
Using these parameters, we start our analysis by evolving the following set of evolution equations: eq.~(\ref{eq:eom_phi_early}), (\ref{eq:ee_hu})-(\ref{eq:ee_hdt}), (\ref{eq:ee_nu})-(\ref{eq:ee_r}), (\ref{eq:ee_nl}) and
\begin{align}
  \dot{a} = aH \,,
\end{align}
where the Hubble parameter, $H$, is given by eq.~(\ref{eq:h}).  This set of evolution equations is solved using eighth-order Runge-Kutta method.

Since the condition for the non-perturbative creation of the Higgses, $q\gg1$ in eq.~(\ref{eq:q_cond}), is dependent on the oscillation amplitude of the inflaton, we have to accurately identify the amplitude of the inflaton.  When the velocity of the inflaton, $\dot{\phi}$, changes sign, we interpolate the values of $\dot{\phi}$ from the previous zero-values using a cubic spline to determine the time when $\dot{\phi}(t^0) = 0$.  Then we interpolate and shift all other outputs from the set of evolution equations back to time $t^0$ and compute $q$ using the amplitude, $\phi(t^0)$.  If $q \leq 1/3$, we stop the non-perturbative creation of Higgses from the inflaton and enter the purely perturbative regime.

Similarly, the determination of the zero-crossing of $\phi$ is just as important because Higgses are created non-perturbatively at $\phi=0$.  Hence, the steps described above are repeated when $\phi$ changes sign.  We then manually increase the number of Higgses by eq.~(\ref{eq:nh0}) and (\ref{eq:nht0}) and decrease the speed of the inflaton by eq.~(\ref{eq:subtract_dphi}) to take into account the instantaneous energy loss due to the creation of the Higgses.

\subsection{Perturbative regime}
\label{sec:p_regime}
Although the inflaton can no longer create Higgses non-perturbatively, we still have to consider the oscillation of the inflaton because the Higgs mass depends on the value of $\phi$.  However, we are no longer interested in the precise time of zero-crossing of the inflaton.  Hence, to simplify numerical calculation, we assume that the inflaton oscillates sinusoidally and convert the inflaton equation of motion to a first order differential equation of the inflaton energy density\footnote{We have used the time averaged result that $\dot{\phi}^2=\rho_\phi$.}:
\begin{align}
  \dot{\rho}_\phi + 3H\rho_\phi + 2\Gamma_{\phi\to h}\rho_\phi + \Gamma_{\phi\to\tilde{h}}\rho_\phi = 0 \,.
  \label{eq:rho_phi_avg}
\end{align}
The matching condition for the non-perturbative and perturbative regimes is $\rho_\phi = m_\phi^2\phi^2_\text{amp}/2$, where $\phi_\text{amp}$ is the value of the inflaton amplitude when the non-perturbative evolution ends.

With this approximation, we convert all quantities that depends on $\phi$ to the corresponding averaged value.  For example, the Higgs mass, which is varying between $0$ to $\alpha\phi_\text{amp}$ becomes
\begin{align}
  \langle m_h \rangle
  = \frac{2}{\pi}m_h^\text{max}
  = \frac{2}{\pi}\alpha\frac{\sqrt{2\rho_\phi}}{m_\phi} \,.
\end{align}
Similarly, all the decay rates are also converted to the appropriate average decay rate, which is shown explicitly in App.~\ref{app:ave}.

In this regime, the following set of evolution equations is solved: eq.~(\ref{eq:rho_phi_avg}), (\ref{eq:ee_hu})-(\ref{eq:ee_hdt}) with $\dot{\phi}^2$ replaced by $\rho_\phi$, (\ref{eq:ee_nu})-(\ref{eq:ee_r}), and (\ref{eq:ee_nl}).  The calculation is continued until the inflaton and the Higgs number densities are smaller than $1\%$ of the number density of the lepton asymmetry.  After this point, the effect of inflaton and Higgses on the lepton asymmetry is insignificant and the inflaton and the Higgs evolution equations are removed from the set of evolution equations to further simplify the calculation.  The set of relevant evolution equations now reduces to eq.~(\ref{eq:ee_sigma}) and
\begin{align}
  &\begin{aligned}
    \dot{n}_{\bar{\nu}_i} + 3Hn_{\bar{\nu}_i}
    &= - \gamma^{-1}_{\bar{\nu}_i}\Gamma_{\bar{\nu}_i\to h_u^\dagger       }n_{\bar{\nu}_i}
       - \gamma^{-1}_{\bar{\nu}_i}\Gamma_{\bar{\nu}_i\to\tilde{h}_u^\dagger}n_{\bar{\nu}_i}
       + 2\Gamma_{\sigma\to\bar{\nu}_i}\frac{\rho_\sigma}{m_\sigma}
    \\\dot{n}_{\tilde{\bar{\nu}}_i} + 3Hn_{\tilde{\bar{\nu}}_i}
    &= - \gamma^{-1}_{\bar{\nu}_i}\Gamma_{\tilde{\bar{\nu}}_i\to h_u^\dagger       }n_{\tilde{\bar{\nu}}_i}
       - \gamma^{-1}_{\bar{\nu}_i}\Gamma_{\tilde{\bar{\nu}}_i\to h_d               }n_{\tilde{\bar{\nu}}_i}
       - \gamma^{-1}_{\bar{\nu}_i}\Gamma_{\tilde{\bar{\nu}}_i\to\tilde{h}_u^\dagger}n_{\tilde{\bar{\nu}}_i}
       + 2\Gamma_{\sigma\to\bar{\nu}_i}\frac{\rho_\sigma}{m_\sigma}
    \\\dot{\rho}_R + 4H\rho_R
    &= \sum_{i=1}^3
       + \gamma^{-1}_{\bar{\nu}_i}\Gamma_{\bar{\nu}_i        \to h_u^\dagger       }M_{R_i}n_{\bar{\nu}_i}
       + \gamma^{-1}_{\bar{\nu}_i}\Gamma_{\bar{\nu}_i        \to\tilde{h}_u^\dagger}M_{R_i}n_{\bar{\nu}_i}
    \\&\,\hspace{1cm} + \gamma^{-1}_{\bar{\nu}_i}\Gamma_{\tilde{\bar{\nu}}_i\to h_u^\dagger       }M_{R_i}n_{\tilde{\bar{\nu}}_i}
       + \gamma^{-1}_{\bar{\nu}_i}\Gamma_{\tilde{\bar{\nu}}_i\to\tilde{h}_u^\dagger}M_{R_i}n_{\tilde{\bar{\nu}}_i}
    \\\dot{n}_L + 3Hn_L
    &= \sum_{i=1}^3
       + \gamma^{-1}_{\bar{\nu}_i}\epsilon_{\bar{\nu}_i}\Gamma_{\bar{\nu}_i        \to h_u^\dagger       }n_{\bar{\nu}_i}
       + \gamma^{-1}_{\bar{\nu}_i}\epsilon_{\bar{\nu}_i}\Gamma_{\bar{\nu}_i        \to\tilde{h}_u^\dagger}n_{\bar{\nu}_i}
    \\&\,\hspace{1cm} + \gamma^{-1}_{\bar{\nu}_i}\epsilon_{\bar{\nu}_i}\Gamma_{\tilde{\bar{\nu}}_i\to h_u^\dagger       }n_{\tilde{\bar{\nu}}_i}
       + \gamma^{-1}_{\bar{\nu}_i}\epsilon_{\bar{\nu}_i}\Gamma_{\tilde{\bar{\nu}}_i\to\tilde{h}_u^\dagger}n_{\tilde{\bar{\nu}}_i} \,.
  \end{aligned}
\end{align}
Notice that the right-handed sneutrinos can no longer decay to the down-type Higgses because the inflaton has decayed out of the system and all remaining terms in the $F$-term of the Higgses are quartic sfermion terms (see eq.~(\ref{eq:f_hu}) and (\ref{eq:f_hd})).\footnote{Note, this is because $m_h=\alpha\langle\phi\rangle=0$ and we neglect three body decays of the sneutrinos.}

This set of evolution equations is evolved until the two heavier right-handed (s)neutrinos decayed and the asymmetry contribution of the lightest right-handed (s)neutrino is less than $1\%$.  Finally, the baryon-to-entropy ratio is calculated from the lepton asymmetry~\cite{Harvey:1990qw}
\begin{align}
  \frac{n_B}{s} = -\frac{8}{23}\frac{n_L}{s} \,.
\end{align}

\section{Results and Discussions}
\label{sec:results}

\subsection{Understanding the Hartree Approximation}
\label{ssec:hartree}
\begin{figure}
  \begin{center}
    \includegraphics[width=0.8\textwidth]{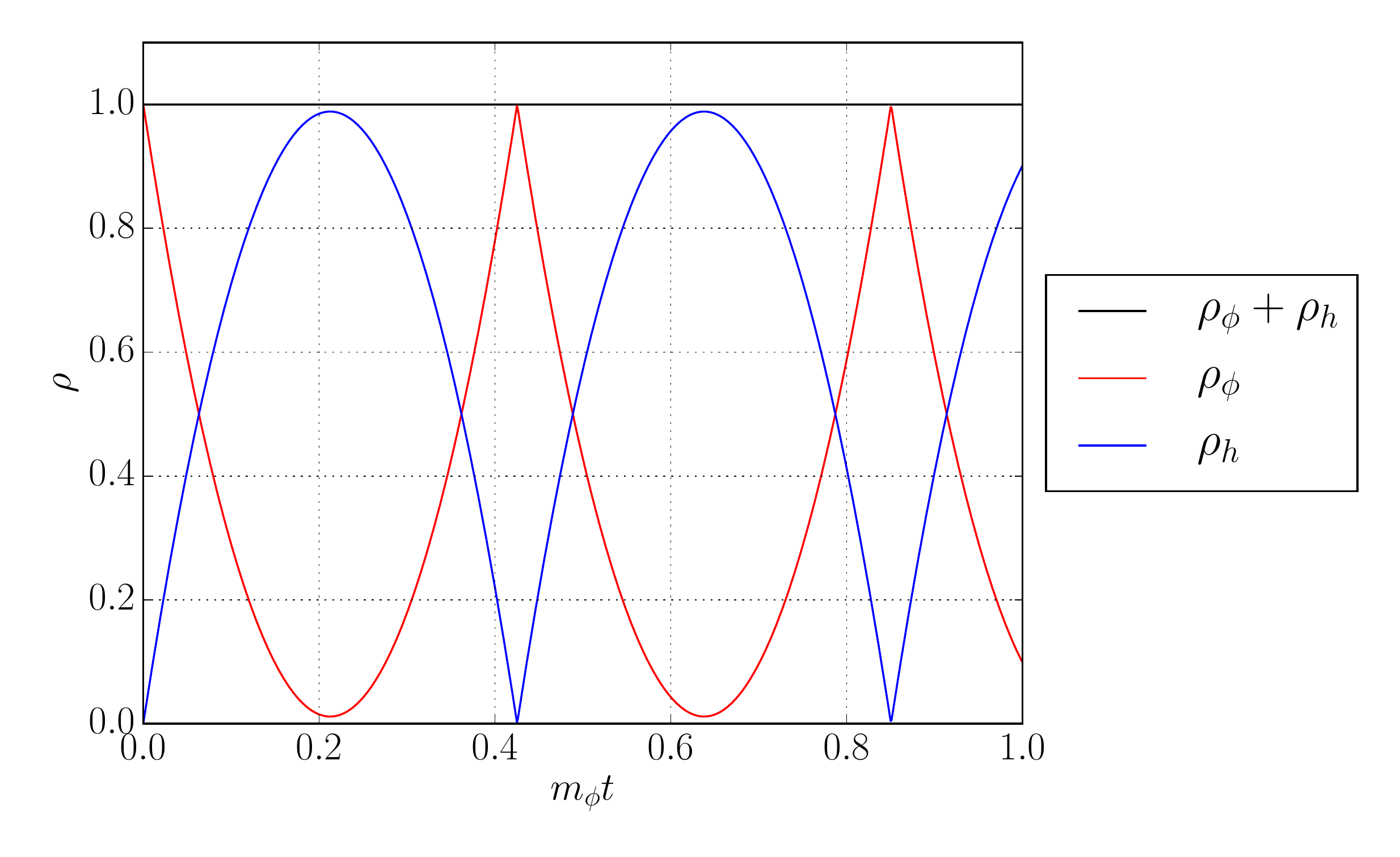}
    \caption{Hartree approximation for $\alpha=1$ in describing the transfer of energy between the inflaton and the up-type Higgs as the inflaton oscillates around its minimum.}
    \label{fig:hartree}
  \end{center}
\end{figure}

To better understand the Hartee approximation, Ref.~\cite{Kofman:1997yn}, consider a simple case with only the up-type Higgs and the inflaton.  In addition, we assume that the non-perturbative creation of Higgses occurs only at the first zero-crossing, and the inflaton and Higgses do not decay perturbatively.  We also ignore the expansion of the universe.  With these assumptions, the relevant evolution equations are
\begin{align}
  \begin{gathered}
    \ddot{\phi} + m_\phi^2\phi + \alpha n_{h_u}\text{sign}(\phi) = 0
    \\\dot{n}_{h_u} = 0 \,.
  \end{gathered}
\end{align}
The solution of this set of evolution equations with $\alpha=1$ is plotted in Fig.~\ref{fig:hartree}, which shows the energy density of the inflaton and that of the up-type Higgs as a function of time.  The energy densities are normalized to the initial energy density in the inflaton field.  From this figure, we see that the up-type Higgses are created with zero mass and do not contribute to the initial energy density.  As the inflaton rolls up the potential, its energy density is transferred to the up-type Higgs while the total energy density of the inflaton-Higgs system stays constant.  Similarly, Higgs energy density is transferred back to the inflaton as the inflaton rolls down the potential.  This simple case shows that the Higgs number density term in the inflaton equation of motion describes the transfer of energy between the inflaton and the Higgs as the inflaton oscillates.

\subsection{Evolving the Evolution Equations}
\begin{figure}
  \begin{center}
    \includegraphics[width=0.8\textwidth]{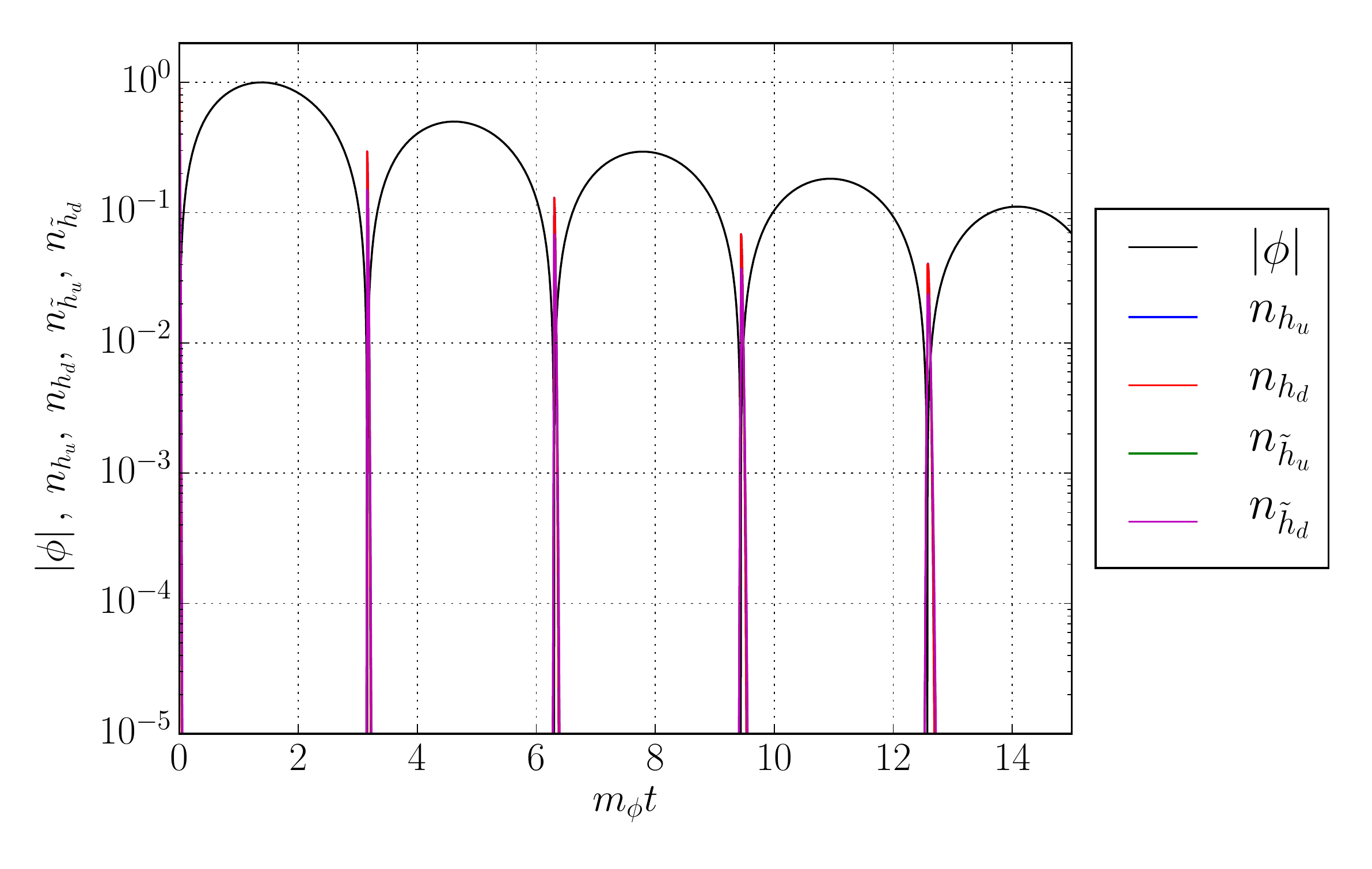}
    \caption{
    The inflaton value in this plot is normalized to the first oscillation amplitude, $2.74\times10^{-1}\,\text{M}_\text{pl}$, while the Higgs number densities are normalized to the up-type Higgs number densities created at the first zero-crossing, $7.23\times10^{-11}\,\text{M}_\text{pl}^3$.  At every zero-crossing of the inflaton, Higgses are created and subsequently decay out of the system almost instantaneously.
    }
    \label{fig:early}
  \end{center}
\end{figure}

At early times, the non-perturbative creation of the Higgses is very efficient.  When the number densities of the created Higgses exceed its thermal equilibrium number density, the Higgses decay to radiation and to the right-handed (s)neutrinos.  This occurs almost instantaneously because the oscillation amplitude of the inflaton is of order Planck scale while the Higgs decay rate is proportional to the Higgs mass, which is proportional to the inflaton vev.  This effect is shown in Fig.~\ref{fig:early}, which shows the magnitude of the inflaton oscillations and the Higgs number densities as a function of time for $\alpha=1$.  The magnitude of the inflaton oscillations is normalized to the first oscillation amplitude, $2.74\times10^{-1}\,\text{M}_\text{pl}$, while the Higgs number densities are normalized to the number of the up-type Higgses created at the first zero-crossing, $7.23\times10^{-1}\,\text{M}_\text{pl}^3$.

\begin{figure}
  \begin{center}
    \includegraphics[width=0.8\textwidth]{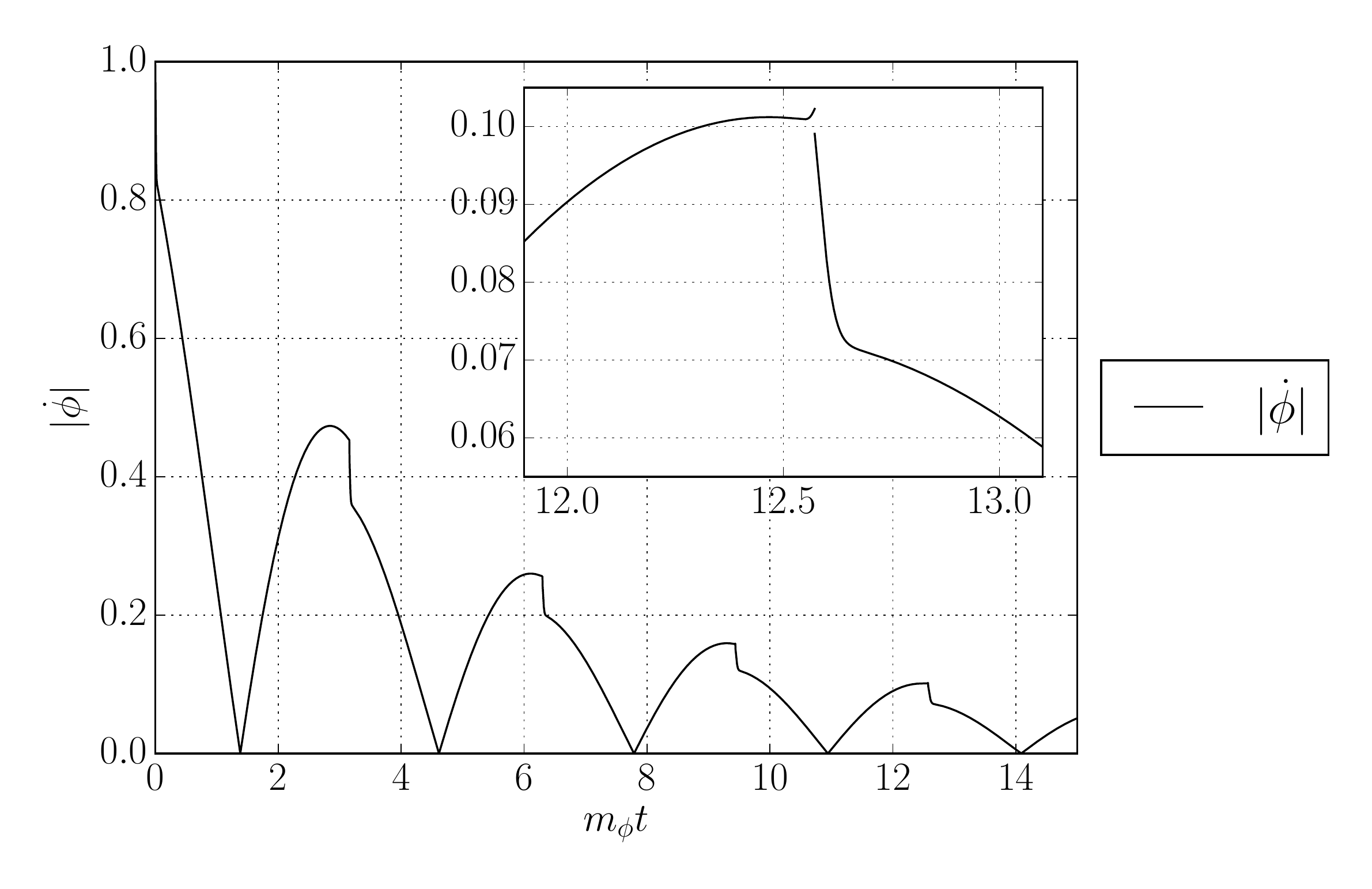}
    \caption{
    The inflaton speed in this plot is normalized to the initial inflaton speed, $2.63\times10^{-6}\,\text{M}_{\text{pl}}^2$.  Since the non-perturbative creation of the Higgs is instantaneous and the subsequent Higgs decay are almost instantaneous, the inflaton experiences a drastic decrease in speed at every zero-crossing.  The Higgs decays transfers energy out from the inflaton-Higgs system.
    }
    \label{fig:early_dphi}
  \end{center}
\end{figure}

Fig.~\ref{fig:early_dphi} shows the inflaton oscillation speed as a function of time for $\alpha=1$.  From this plot, we see that the inflaton experiences a drastic decrease in speed at every zero-crossing due to the almost instantaneous decay of the Higgses.  When the Higgses decay, energy is transferred out from the inflaton-Higgs system to the decay products.  In addition, this drastic decrease does not occur when the inflaton speed is at its maximum because the inflaton oscillation is damped.  The sub-figure, which is the zoomed-in version of the plot, shows the typical behavior of the inflaton speed at zero-crossing.  As shown in the sub-figure, the inflaton speed has a discontinuous drop, which is due to the energy lost in the non-perturbative creation of the Higgses.  The sub-figure also shows that the inflaton speed increases momentarily before the zero-crossing.  As the inflaton rolls down the potential, the Higgs mass decreases.  Once the Higgses become lighter than the right-handed (s)neutrinos, the right-handed (s)neutrinos start to decay to the Higgses, increasing the energy in the inflaton-Higgs system.  However, as shown in Sec.~\ref{ssec:hartree}, as the inflaton rolls down the potential, energy is transferred from the Higgses to the inflaton.  Hence, the inflaton speed increases for a short period of time before reaching the bottom of the potential.

The dynamics in the non-perturbative regime with $q \gg 1$ is dominated by a broad parametric resonance.  When $q \leq 1/3$, and the inflaton continues to oscillate, there is only a narrow region in momentum space where parametric resonance can occur.  At every passage of $\phi$ through zero, Higgses are produced, however they quickly thermalize on an oscillation time scale.  Thus these Higgses acquire momentum outside the region of the parametric resonance, which in effect suppresses the additive effect.  It is at this point that instant preheating ceases to dominate.

\begin{figure}
  \begin{center}
    \includegraphics[width=0.8\textwidth]{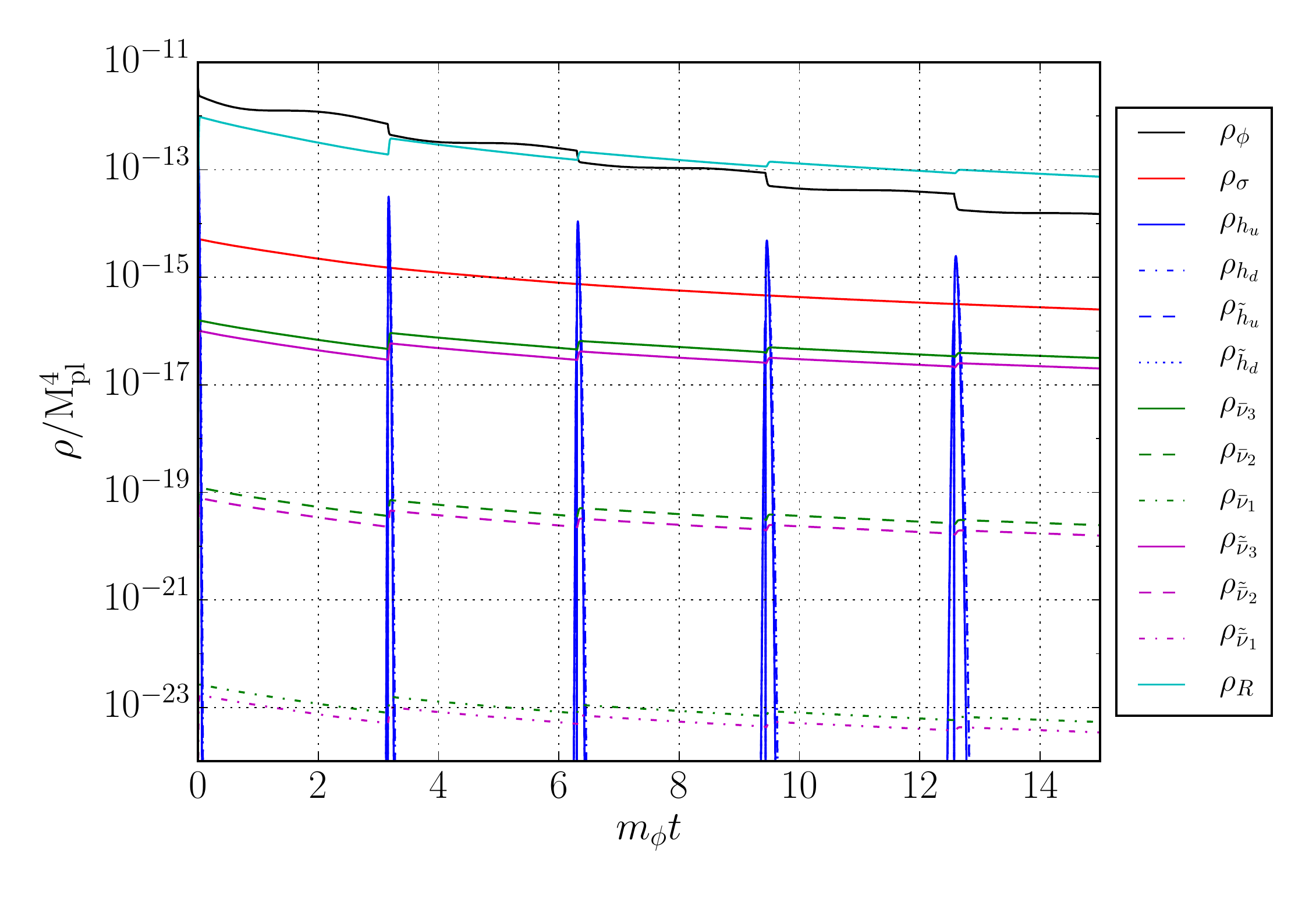}
    \caption{
    Since the universe is radiation dominated after a couple inflaton oscillations, the reheating process of our model is very efficient.
    }
    \label{fig:early_rho}
  \end{center}
\end{figure}

Fig.~\ref{fig:early_rho} shows the energy densities of the inflaton, the waterfall fields, radiation, and the right-handed (s)neutrinos as a function of time for $\alpha=1$.  From this figure, we see that after a couple of inflaton oscillations, radiation energy density dominates over all other energy densities.  The associated reheat temperature is of order $T_\text{reheat}\sim10^{15}\,\text{GeV}$, i.e.~less than the GUT scale.\footnote{This will have consequences for gravitino bounds which we discuss in the conclusion.}  This shows that the reheating process of our model is very efficient.  The epoch of radiation domination occurs later as $\alpha$ decreases because the non-perturbative and the perturbative decay rates of the inflaton decrease as $\alpha$ decreases.  In addition, we clearly see from this figure that the energy densities of the inflaton and the right-handed (s)neutrinos increase or decrease like a step-like function.  This step-like function behavior is due to the non-perturbative creation of the Higgs that only occurs at zero-crossing.

\begin{figure}
  \begin{center}
    \includegraphics[width=0.8\textwidth]{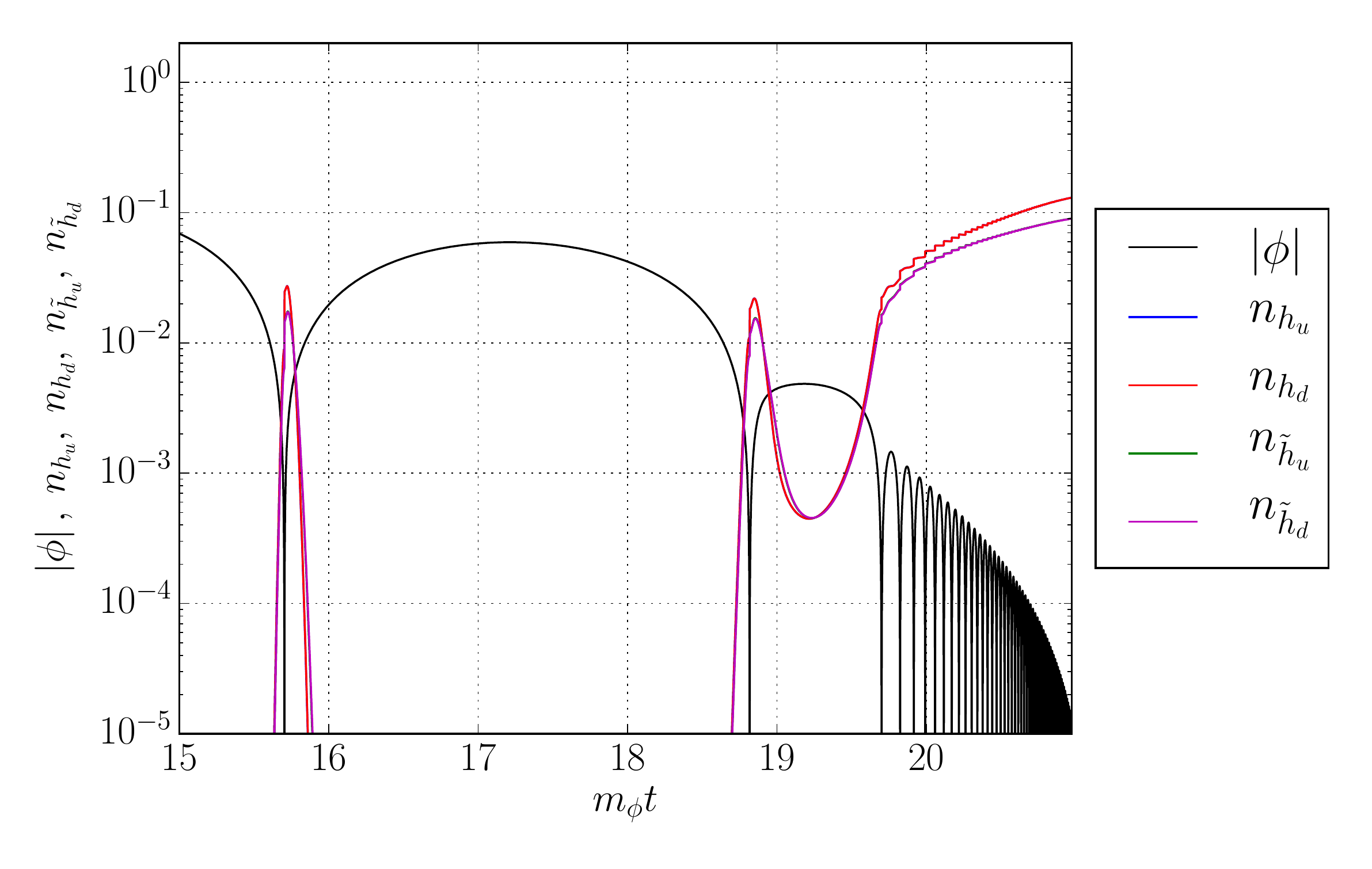}
    \caption{
    This plot is a continuation of Fig.~\ref{fig:early}.  The decrease in inflaton oscillation amplitude increases the Higgs number densities by decreasing the Higgs decay rate.  The increase in the Higgs number densities increases the inflaton oscillation frequency, which further decreases the inflaton oscillation amplitude by increasing the frequency of inflaton non-perturbative creation of the Higgs.
    }
    \label{fig:phi_nh_b}
  \end{center}
\end{figure}

As the inflaton oscillation amplitude decreases, the Higgs decay occurs slower, which increases the Higgs number densities in the system (see Fig.~\ref{fig:phi_nh_b}).  This effect increases the inflaton oscillation frequency because, from the Hartree approximation in eq.~(\ref{eq:eom_phi_early}), the oscillation frequency is given by
\begin{align}
  \omega = m_\phi^2+\alpha^2\langle h^2\rangle = m_\phi^2+\alpha\frac{n_h}{|\phi|} \,.
  \label{eq:oscfreq}
\end{align}
The increase in the inflaton oscillation frequency further increases the Higgs number densities because the non-perturbative creation now occurs more frequently.  In addition, the increase in the Higgs number densities decreases the inflaton oscillation amplitude because the Higgses are taking away more energy from the inflaton.  Notice that this effect produces a feedback effect that decreases the amplitude of the inflaton oscillation at a faster rate.  This effect can be seen in Fig.~\ref{fig:phi_nh_b}, which shows the inflaton oscillation amplitude and the Higgs number densities as a function of time with $\alpha=1$.  This figure is the continuation of Fig.~\ref{fig:early}.

\subsection{Baryon Asymmetry}
The final lepton asymmetry number density can be converted to that of a baryon asymmetry via the sphaleron process, since the sphaleron process violates $n_{B+L}$ but conserves $n_{B-L}$.  The conversion factor between the baryon and lepton asymmetries in the MSSM is given by~\cite{Harvey:1990qw}
\begin{align}
  n_B = -\frac{8}{23}n_L \,.
\end{align}
Thus, the baryon-to-entropy ratio can be obtained by
\begin{align}
  \frac{n_B}{s} = -\frac{8}{23}\frac{n_L}{s} \,,
\end{align}
where the entropy can be calculated from the radiation energy density using
\begin{align}
  \begin{aligned}
    \rho_R &= \frac{\pi^2}{90}gT^4
    \\s &= \frac{2\pi^2}{45}gT^3 \,.
  \end{aligned}
\end{align}
The experimentally measured value for the baryon-to-entropy ratio is~\cite{Ade:2015xua}
\begin{align}
  \frac{n_B}{s} \sim 8.7\times10^{-11} \,.
\end{align}

By fitting to the low energy observables, the asymmetry produced by the heaviest right-handed (s)neutrinos has the correct sign while that produced by the two lighter ones have the opposite sign as shown in eq.~(\ref{eq:parameters}).  Hence, to reproduce the measured baryon-to-entropy ratio, we need to strike a balance between the number density of the heaviest right-handed (s)neutrinos and that of the two lighter ones.  Luckily, this is achievable in our model because the number density of Higgses created non-perturbatively, the Higgs decay rates and the Higgs mass are controlled by the inflaton-Higgs coupling, $\alpha$, which is a free parameter.

\begin{figure}
  \begin{center}
    \includegraphics[height=0.3\textheight]{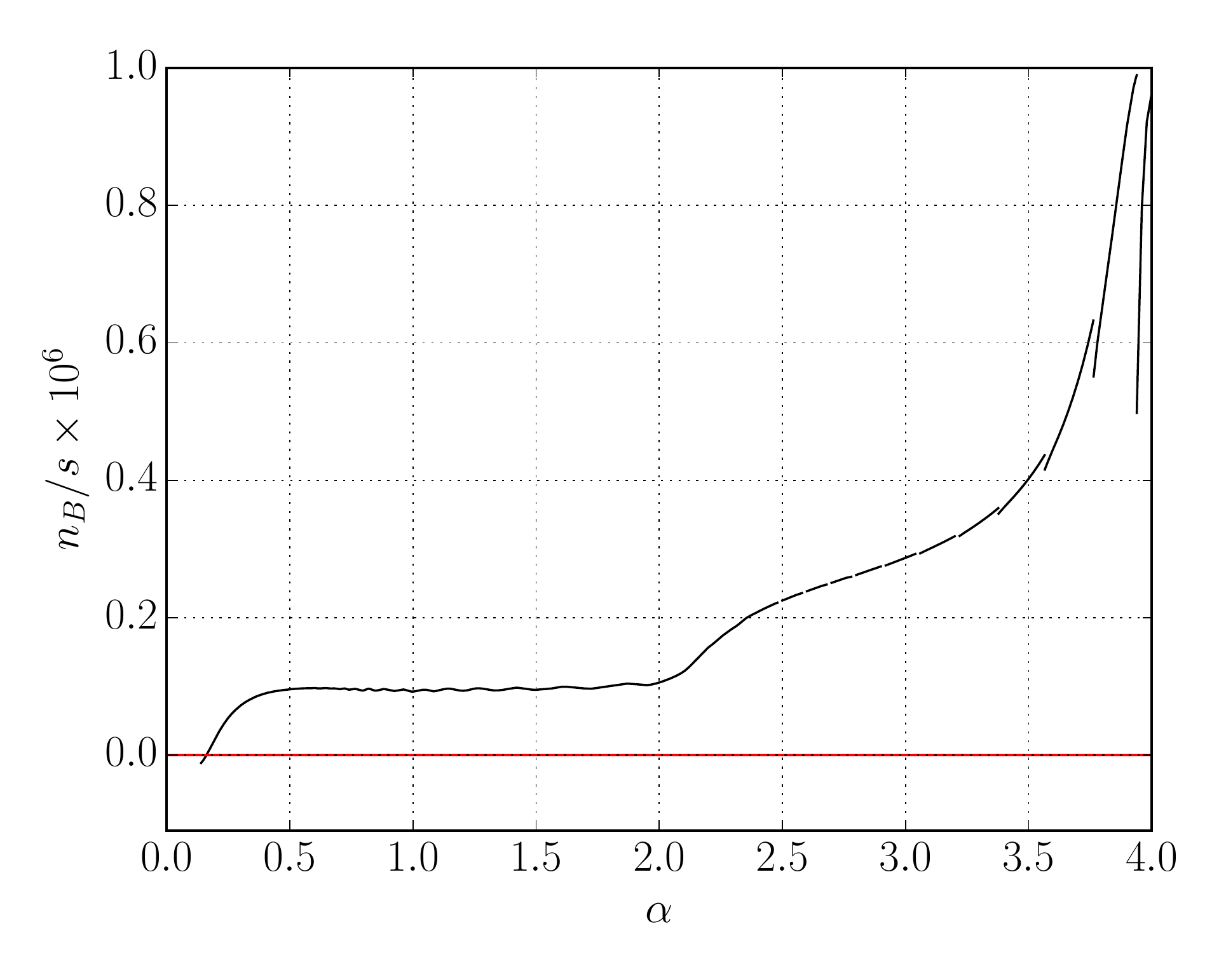}
    \includegraphics[height=0.3\textheight]{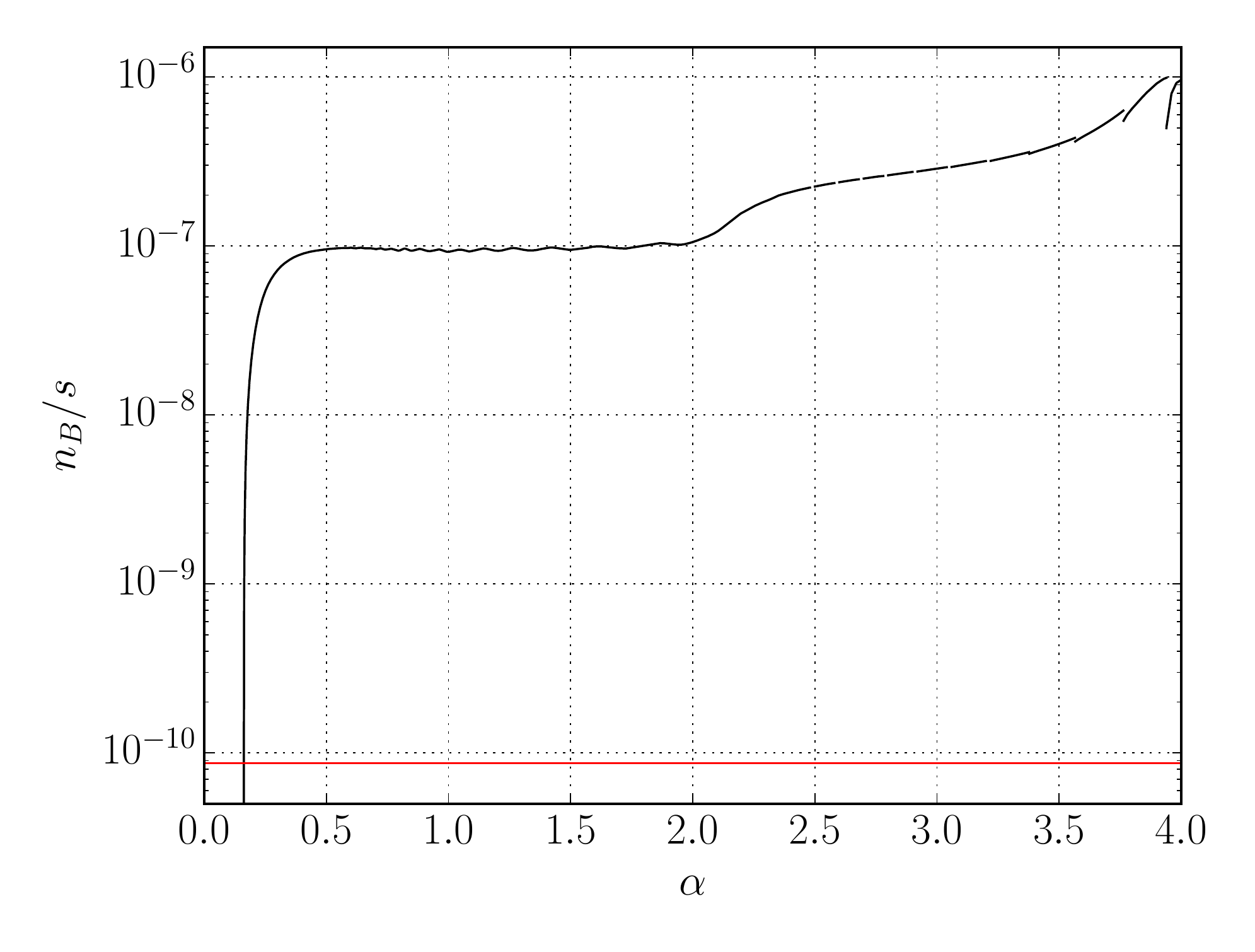}

    \includegraphics[height=0.3\textheight]{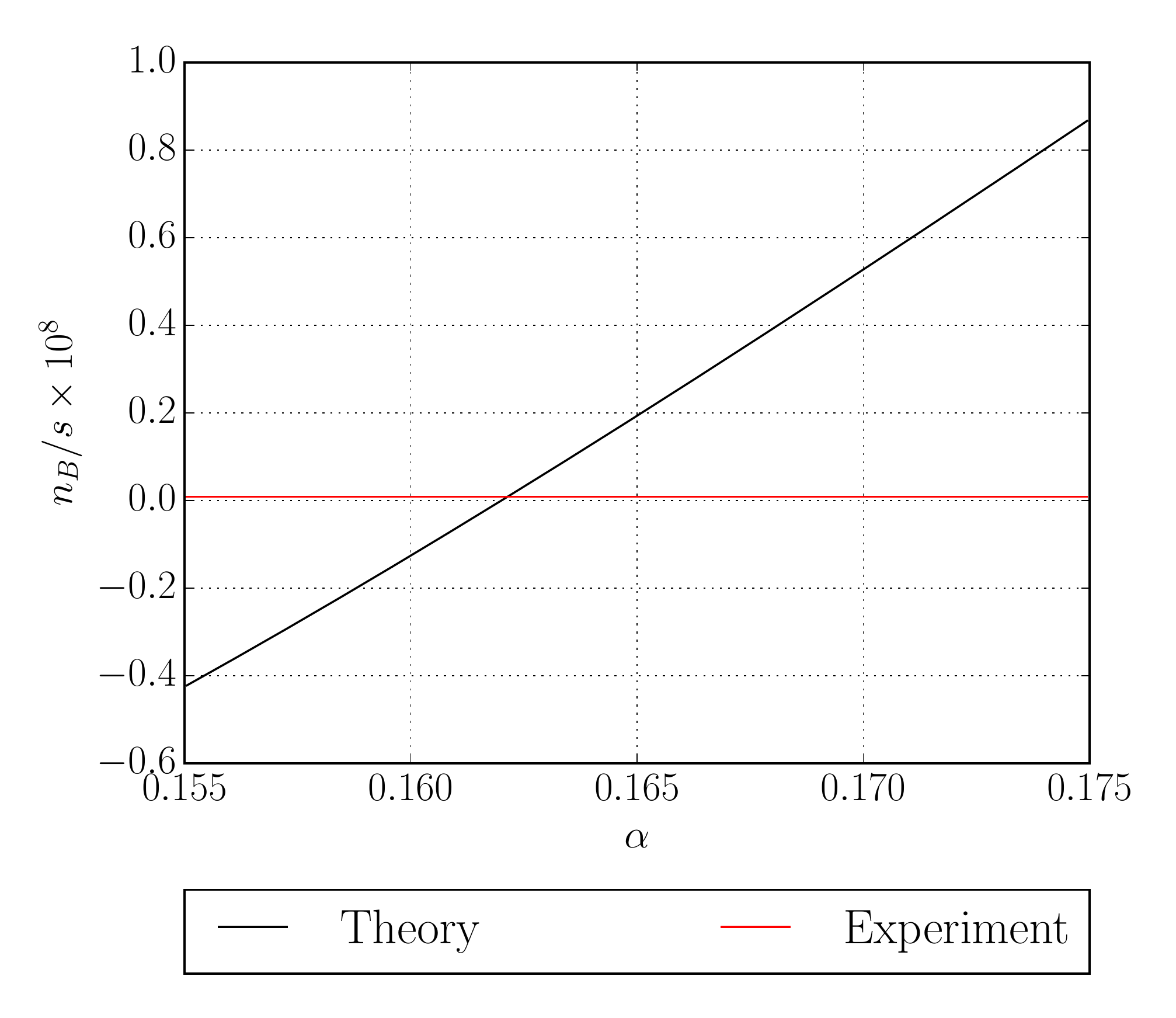}
    \caption{
    All three figures show the baryon-to-entropy ratio as a function of $\alpha$.  The top figure is in a linear-scale, the middle figure has a log-scale in the $y$-axis, while the $x$-axis of the bottom figure is zoomed-in with linear scale.  The baryon-to-entropy ratio matches the observed value for $\alpha\sim0.162$.
    }
    \label{fig:nb_v_alpha}
  \end{center}
\end{figure}

Fig.~\ref{fig:nb_v_alpha} shows the baryon-to-entropy ratio as a function of $\alpha$.  The top figure is in a linear scale, the second figure has a log-scale $y$-axis, while the bottom figure is the zoomed in version of the top figure for a value of $\alpha$ that fits the observed baryon-to-entropy ratio.  This figure has some very curious features that require some explanations. 1) Ignoring the discontinuity, the baryon-to-entropy ratio decreases as $\alpha$ decreases. 2) The baryon-to-entropy ratio is not a continuous function of $\alpha$. 3) As $\alpha$ decreases, the discontinuity in the baryon-to-entropy ratio decreases. 4) The baryon-to-entropy ratio plateaus at around $\sim10^{-7}$ for a wide range of $\alpha$. We will address each of these features one by one in the following paragraphs.

\begin{figure}
  \begin{center}
    \includegraphics[width=0.7\textwidth]{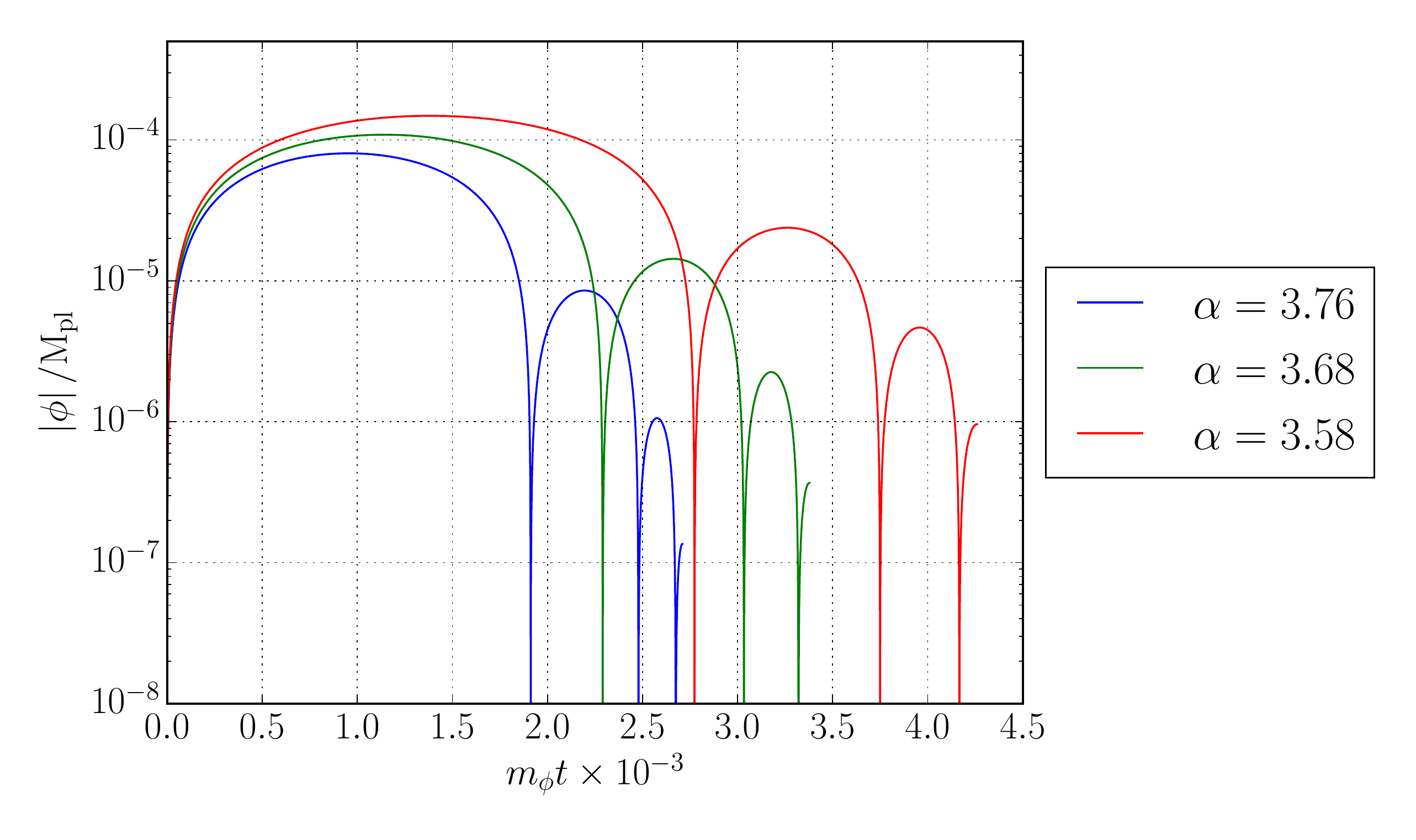}
    \includegraphics[width=0.7\textwidth]{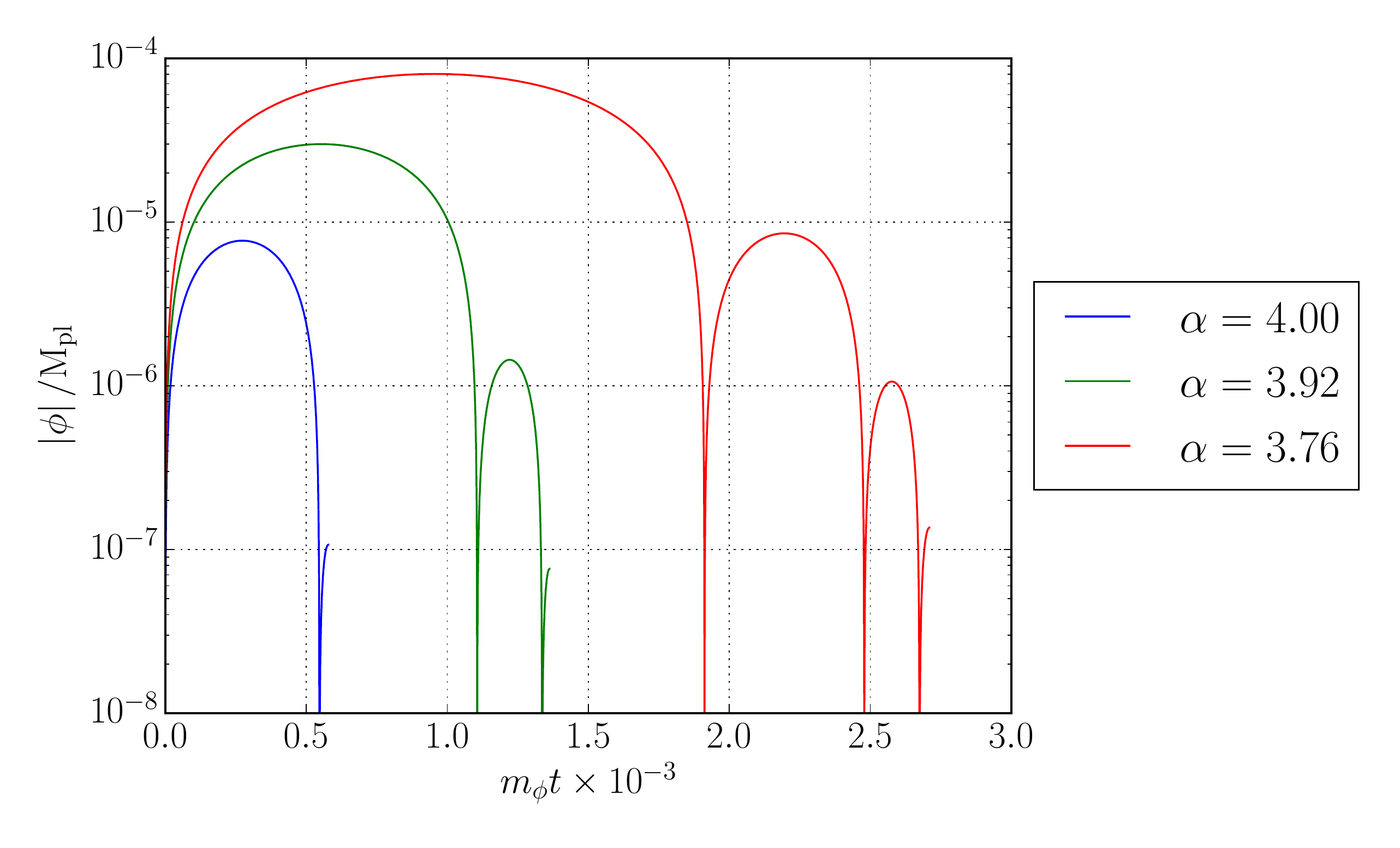}
    \caption{
      These figures show inflaton oscillations in the non-perturbative regime for various values of $\alpha$.  The plot ends when the system exits the non-perturbative regime.  The top figure shows $\alpha$ that are on the same branch in Fig.~\ref{fig:nb_v_alpha}, while the bottom figure shows $\alpha$ across different branches.  From the top figure, we see that as $\alpha$ decreases the inflaton energy density when the system exits the non-perturbative regime increases.  In the perturbative regime, the Higgs can only decay to the two lighter right-handed (s)neutrinos.  Hence, as $\alpha$ decreases, the baryon-to-entropy ratio decreases.  On the other hand, the bottom figure shows that as $\alpha$ decreases, the number of inflaton oscillation in non-perturbative regime increases causing the abundance ratio for the heaviest right-handed (s)neutrinos to the two lighter ones to increase discontinuously.  This is the root cause for the discontinuity in Fig.~\ref{fig:nb_v_alpha}.
    }
    \label{fig:nb_a_explain}
  \end{center}
\end{figure}

The decrease in the baryon-to-entropy ratio as $\alpha$ decreases can be attributed to two factors.  First, in the non-perturbative regime, the inflaton energy density decreases mainly due the non-perturbative creation of Higgses at zero-crossing.  Since the number density of Higgs created at zero-crossing is proportional to $\alpha^{3/2}$ (see eq.~\ref{eq:n_chi_created}), the inflaton energy density when the system exits the non-perturbative regime is larger for smaller $\alpha$.  Recall that in the non-perturbative regime, the Higgses decay predominantly to the heaviest right-handed (s)neutrinos because of the larger Yukawa coupling.  However, in the perturbative regime $q\lesssim 1/3$ (see eq.~\ref{eq:q_cond}), which means
\begin{align}
  m_h^\text{max} \lesssim 10^{-6}\,\text{M}_\text{pl} < M_{R_3} \sim 10^{-5}\,\text{M}_\text{pl} \,,
\end{align}
and the Higgses can decay only to the two lighter right-handed (s)neutrinos.  Hence, the inflaton energy density when the system exits the non-perturbative regime will eventually be transferred to the two lighter right-handed (s)neutrinos and radiation decreasing the baryon-to-entropy ratio.  This effect is shown in the top figure of Fig.~\ref{fig:nb_a_explain}, where the magnitude of the inflaton oscillation amplitude is plotted as a function of time.  The figure ends when the system exits the non-perturbative regime.  As shown in the figure the inflaton oscillation amplitude is larger when the system exits the non-perturbative regime for smaller $\alpha$.

Second, the decay of Higgses to the right-handed (s)neutrinos and the decay of the right-handed (s)neutrinos to the Higgses both contribute to the lepton asymmetry.  Due to the inflaton oscillation, the Higgses decays to the right-handed (s)neutrinos when inflaton is close to the oscillation amplitude, while the right-handed (s)neutrinos decay to the Higgses when the inflaton is near the bottom of the potential.  Hence, the amount of asymmetry created increases as the number of inflaton oscillation increases.  Since the inflaton perturbative decay rates are proportional to $\alpha^2$ (see eq.~\ref{eq:phi2h}), the inflaton goes through more oscillations before decaying out of the system for smaller $\alpha$.  Given that the two lighter right-handed (s)neutrinos contribute to the baryon-to-entropy ratio with the wrong sign, the baryon-to-entropy ratio decreases as $\alpha$ decreases.

The discontinuity in the baryon-to-entropy ratio is best understood by considering the abundance ratio for the heaviest right-handed (s)neutrino to the two lighter ones.  When this abundance ratio is larger, the baryon-to-entropy ratio is more positive because the heaviest right-handed (s)neutrino decays contribute to the asymmetry with the correct sign.  In the non-perturbative regime, Higgses are created at every zero-crossing and subsequently decay predominantly to the heaviest right-handed (s)neutrinos.  Hence, the abundance ratio increases discontinuously at every zero-crossing.  The bottom figure of Fig.~\ref{fig:nb_a_explain} shows the magnitude of the inflaton oscillation as a function time.  The plot ends when the system exits the non-perturbative regime.  As shown in the figure, the inflaton goes through more oscillations as $\alpha$ decreases because the number density of Higgses created at zero-crossing is proportional to $\alpha^{3/2}$ (see eq.~\ref{eq:n_chi_created}).  Hence, the inflaton oscillation amplitude decreases slower for smaller $\alpha$, thus the broad parametric resonance parameter, $q\propto\phi_\text{amp}^2$ decreases slower.  Along with Fig.~\ref{fig:nb_v_alpha}, we see that as the number of inflaton oscillations in the non-perturbative regime increases, the baryon-to-entropy ratio increases.  In addition, all three $\alpha$ in the top figure of Fig.~\ref{fig:nb_a_explain} are within the same branch of baryon-to-entropy ratio in Fig.~\ref{fig:nb_v_alpha}.  This shows that the discontinuous jump only occurs when the total number of inflaton oscillations in the non-perturbative regime increases.

The discontinuity in baryon-to-entropy ratio decreases as $\alpha$ decreases because as $\alpha$ decreases the number density of Higgses created at zero-crossing decreases.  Therefore, the contribution to the heaviest right-handed (s)neutrinos by each additional inflaton oscillation in the non-perturbative regime decreases as $\alpha$ decreases.  Hence, the discontinuity of the ratio of the heaviest right-handed (s)neutrinos to the two lighter ones decreases as $\alpha$ decreases.  This ratio is directly related to the baryon-to-entropy ratio, resulting in a smaller discontinuity in baryon-to-entropy ratio as $\alpha$ decreases.

From these explanations, we see that majority of the factors contribute negatively to the baryon-to-entropy ratio as $\alpha$ decreases, while the increase in the number of inflaton oscillations, in the non-perturbative regime, increases the baryon-to-entropy ratio as $\alpha$ decreases.  The plateau of baryon-to-entropy ratio is created when the we have a balance between these factors.

In Fig.~\ref{fig:nb_v_alpha}, we plotted the baryon-to-entropy ratio for $\alpha\lesssim4$ because from eq.~(\ref{eq:subtract_dphi}), we see that the upper limit for $\alpha$ is given by
\begin{align}
  \alpha^\text{max} = \frac{\pi^2}{\sqrt{6}} \sim 4.03 \,.
\end{align}
With this value of $\alpha$, all energy in the inflaton is transferred to the Higgses at the first zero-crossing.  Since there are no inflatons leftover to oscillate, the Higgses are massless and they never decay to the right-handed (s)neutrinos.  Hence, the only source of right-handed (s)neutrinos comes from the waterfall field.  In this situation, we do not obtain the observed baryon-to-entropy ratio because the waterfall field decays predominantly to the heaviest right-handed (s)neutrinos creating an over abundance of baryons.

To summarize, the inflaton-Higgs coupling, $\alpha$, which is a free parameter in our model, can be tuned to reproduce the observed baryon-to-entropy ratio.  A more natural model would, however, have the plateau of the baryon-to-entropy ratio in Fig.~\ref{fig:nb_v_alpha} closer to the experimental value.\footnote{A recent paper by Raymond Co.~et.~al.  \cite{Co:2016fln}, points out that for models of gravitino or axino dark matter with high reheat temperature, such as our model, the decay of the saxions can produce 3 orders of magnitude of entropy in the case that the saxion vev, $s_I$, equals the PQ breaking vev, $V_{PQ}$ and $s_I = V_{PQ} \approx 10^{14}$ GeV (see Eqn. 2.9, Ref. \cite{Co:2016fln}).  This would make the plateau of our baryon-to-entropy ratio of order the observed experimental value.}

\subsection{Sources of Uncertainties}
It is important to note that there are multiple sources of uncertainties in our calculation.  An obvious and main source of theoretical uncertainty is the neglect of momentum distributions and the exclusion of the $2-2$ scattering from our simulation.  Hence, there are implicit error bars in each of the plots shown in this paper.  An interesting follow-up project would be to include the full Boltzmann equations treatment and the $2-2$ scattering to remove this theoretical uncertainty.

Another interesting source of uncertainty is due to the nature of this system where the masses, the Yukawa couplings, and the CP asymmetry parameter of the heaviest and two lighter right-handed (s)neutrinos are separated by a couple orders of magnitude.  Hence, the number densities of the heaviest and the two lighter right-handed (s)neutrinos differ by orders of magnitude when they are created.  To obtain the observed baryon-to-entropy ratio, our model requires a precise cancellation between the CP asymmetry created by the heaviest right handed (s)neutrinos and that created by the two lighter ones.  Therefore, unless we are able to keep track of the number densities of the heaviest right-handed (s)neutrinos to high accuracy, numerical errors will creep into the calculation.  With this in mind, the plots of the baryon-to-entropy ratio as function of $\alpha$, shown in Fig.~\ref{fig:nb_v_alpha}, is obtained by line fitting through the calculated values.

Finally, we also want to point out that the set of evolution equations are a set of non-linear, highly coupled, and stiff differential equations.  Numerical solutions to differential equations with this characteristic are unstable, unless the time step is taken to be very small.  To prevent this problem, we have checked that reducing the time step only changes the baryon-to-entropy ratio by a negligible amount.

\section{Summary}
This paper is an extension of the previous Pati-Salam subcritical hybrid inflation paper, proposed by two of us~\cite{Bryant:2016tzg}, which was shown to successfully reproduce inflation observables.  In this paper, we studied the reheating process and the baryogenesis via leptogenesis of the model.

In the instant preheating process, the coupling of the inflaton to the Higgses causes the inflaton to non-perturbatively decay to Higgses efficiently as it oscillates around its minimum.  The produced Higgses then decay to radiation and reheat the universe.  The reheat temperature depends on the value of $\alpha$.  We find $T_\text{reheat}\sim10^{15}\,\text{GeV}$.  This can, in principle, create a cosmological problem with gravitinos.  However, for gravitino masses greater than $\sim 40$ TeV, the only problem concerns the over-closure of the universe by an LSP with mass of order 100 GeV~\cite{Kawasaki:2008qe}.  This suggests that the LSP in our model would need to be a light axino in conjunction with an axion dark matter candidate.  This then ties in interestingly to the scenario of Ref. \cite{Co:2016fln}.

As for baryogenesis, fitting to low-energy observables forces the CP asymmetry parameter of the heaviest right-handed (s)neutrinos to have the correct sign, while that of the two lighter right-handed (s)neutrinos have the wrong sign.  Hence, it is important to include all three right-handed (s)neutrinos in our analysis.  In the model, the coupling parameter, $\alpha$, between the inflaton and the Higgses is a free parameter.  By tuning this parameter, we can obtain the observed baryon-to-entropy ratio.

The matter sector of our model has $24$ input parameters.  By fitting our model to $49$ low-energy observables, our model has $25$ degrees of freedom.  We obtain a reasonable fit to the low-energy observables~\cite{Raby:2015rex,Poh:2015wta,Anandakrishnan:2014nea,Raby:2013wva,Anandakrishnan:2013pja,Anandakrishnan:2013nca,Anandakrishnan:2012tj}.  A benchmark point is given in App.~\ref{app:low_energy}.  The $\chi^2/\text{dof}$ of this benchmark point is $\chi^2=1.90$, which corresponds to a $p$-value of $0.004$.\footnote{The worse fits come from the up and down quark masses and $\sin2\beta$.  We have obtained a significantly better fit of $\chi^2/\text{dof}=1.24$, which corresponds to a $p$-value of $0.19$, with the addition of one additional complex parameter in the Yukawa matrices. We have further checked that the CP asymmetry of the three right-handed (s)neutrinos are of the same order of magnitude.  Hence, the results of this paper are not affected.  This will be presented in a future paper~\cite{Poh:new}.} In this benchmark point, the first two family scalars have mass $\sim25\,\text{TeV}$ while the third family scalars have mass $\sim5\,\text{TeV}$.  The scalars in our model do not decouple completely from the low-energy observables.  The heavy Higgs, charged Higgs, and the CP-odd Higgs all have mass around $2\,\text{TeV}$ while the NLSP of our model is a neutralino.  The gluinos in our model have a best fit mass lighter than $\sim 2\,\text{TeV}$, which has the potential to be observed in the current run of the LHC. The dominant experimental signature for the gluino is $b$-jets with leptons and missing $E_T$~\cite{Anandakrishnan:2014nea,Anandakrishnan:2013nca}.  In this paper, we extended the previous studies by describing the reheating and baryogenesis via leptogenesis of this model.

\section*{Acknowledgments}

We are indebted to Radovan Derm\'i\v{s}ek for his program and his valuable inputs in using it.  B.C.B., Z.P.~and S.R.~received partial support for this work from DOE/ DE-SC0011726.   We are also grateful to Gary Steigman and Hong Zhang for discussions.

\appendix
\section{Yukawa Matrices Conventions}
\label{app:yukawas}
In Sec.~\ref{sec:review}, the Yukawa matrices are defined in Weyl notation with doublets on the left.  The Yukawa matrices with this definition are denoted as $Y$.

From Sec.~\ref{sec:broad_stroke} onwards, the Yukawa matrices are defined in Weyl notation with doublets on the right.  The Yukawa matrices with this definition are denoted as $\lambda$.  The reason that we switch to this definition is because the renormalization group equations in the global $\chi^2$ analysis program that we use, \texttt{maton}, is written with doublets on the right.

As a comparison with the literature, the definition of Yukawa matrices in the relevant references regarding CP asymmetry are different.  Here we provide a dictionary.  The Yukawa matrices in Covi et.~al.~\cite{Covi:1996wh}, $\lambda^{CRV}$; in Giudice et.~al.~\cite{Giudice:2003jh}, $Y^{GNRRS}$;  in Buchmuller et.~al.~\cite{Buchmuller:2005eh}, $h^{BPY}$; and in Davidson et.~al.~\cite{Davidson:2008bu} , $\lambda^{DNN}$; are related to our Yukawa matrix, $\lambda^{BPR}$, as follows:
\begin{align}
  \lambda^{BPR} = {\lambda^{CRV}}^T = {Y^{GNRRS}}^T = h^{BPY} = {\lambda^{DNN}}^T \,.
\end{align}

\section{Global $\chi^2$ Fits to Low-energy Data}
\label{app:low_energy}

The parameters of our model are listed in Table~\ref{tab:inputs}.  A detailed global $\chi^2$ analysis of our model can be found in~\cite{Raby:2015rex,Poh:2015wta,Anandakrishnan:2014nea,Raby:2013wva,Anandakrishnan:2013pja,Anandakrishnan:2013nca,Anandakrishnan:2012tj}.  Comparing with~\cite{Poh:2015wta}, we have added the following new observables: $V_{ud},V_{cd},V_{cs},V_{tb}$ and updated all experimental values to the latest values.  The experimental values for neutrino observables are the best fit points in Capozzi et.~al.~while that for $B\to K^*\mu^+\mu^-$ observables are the latest LHCb results~\cite{Capozzi:2016rtj,Aaij:2013iag,Aaij:2013qta}.  All other experimental values are the latest values in Particle Data Group and Heavy Flavor Averaging Group~\cite{Agashe:2014kda,Amhis:2014hma}.  We also updated the public codes, \texttt{superiso} and \texttt{susy flavor}, that are used to calculate flavor observables~\cite{Mahmoudi:2008tp,Rosiek:2014sia}.

The theoretical errors are also increased by a few percent due to the large scatter of SUSY particle masses between $m_{16}$ and $M_Z$.  We run renormalization group equations from the lightest right-handed neutrino mass scale to the electroweak scale using two-loop renormalization group equations and remove the over-running by performing one-loop threshold corrections.  The theoretical errors are obtained by performing threshold corrections in different orders; that is performing gauge threshold corrections before or after Yukawa threshold corrections.  These errors were neglected in previous studies.
\begin{table}[!htbp]
  \begin{center}
    \renewcommand{\arraystretch}{1.2}
      \scalebox{0.83}{
        \begin{tabular}{|l|c|c|}
          \hline Sector & Parameters & No.
          \\ \hline Gauge & $\alpha_G$, $M_G$, $\epsilon_3$ & 3
          \\ SUSY (GUT scale) & $m_{16}$, $M_{1/2}$, $A_0$, $m_{H_u}$, $m_{H_d}$ & 5
          \\ Yukawa Textures & $\epsilon$, $\epsilon'$, $\lambda$, $\rho$, $\sigma$, $\tilde\epsilon$, $\xi$, $\phi_{\rho}$, $\phi_{\sigma}$, $\phi_{\tilde\epsilon}$, $\phi_{\xi}$ & 11
          \\ Neutrino & $M_{R_1}$, $M_{R_2}$, $M_{R_3}$ & 3
          \\ SUSY (EW Scale) & $\tan\beta$, $\mu$ & 2
          \\ \hline Total & & 24
          \\ \hline
        \end{tabular}
      }
    \caption{
    The matter sector has 24 input parameters
    }
    \label{tab:inputs}
  \end{center}
\end{table}

Table~\ref{tab:benchmark} contains the experimental and fit values for the $49$ observables along with the pulls.  All the masses in the table are in GeV. The gluino mass for this model is $M_{\tilde{g}}\sim1.5\,\text{TeV}$.  The $\chi^2/\text{dof}$ of this benchmark point is $\chi^2/\text{dof}=1.90$ and the corresponding $p$-value is $0.004$.

The input parameters to the benchmark point in Table~\ref{tab:benchmark} are
\begin{align*}
  \begin{aligned}
    (1/\alpha_G, M_G, \epsilon_3) &= (26.17, 2.31\times10^{16}\,\text{GeV}, -0.68\%)\,,
    \\(\lambda, \lambda\epsilon, \sigma, \lambda\tilde\epsilon, \rho, \lambda\epsilon', \lambda\epsilon\xi) &= (0.6061, 0.0306, 1.2476, 0.0044, 0.0741, -0.0018, 0.0036)\,,
    \\(\phi_\sigma, \phi_{\tilde\epsilon}, \phi_{\rho}, \phi_\xi) &= (0.50, 0.53, 3.99, 3.46)\text{rad}\,,
    \\(m_{16}, M_{1/2}, A_0, \mu(M_Z)) &= (25000, 400, -51387, 994)\,\text{GeV} \,,
    \\((m_{H_d}/m_{16})^2, (m_{H_u}/m_{16})^2, \tan\beta) &= (1.89, 1.61, 50.34)\,,
    \\(M_{R_1}, M_{R_2}, M_{R_3}) &= (9.0, 573.9, 29532.4)\times10^9\,\text{GeV} \,.
  \end{aligned}
\end{align*}

\begin{table}[!htbp]\footnotesize
  \caption{Benchmark point with $m_{16} = 25\ \text{TeV}, M_{\tilde g} = 1.493\ \text{TeV}$:}
  \label{tab:benchmark}
  \begin{center}
    \scalebox{0.8}{
      \begin{tabular}{|l|r|r|r|r|}
        \hline
        Observable & Fit & Exp. & Pull & $\sigma$ \\
        \hline\hline
        $M_Z$ & 91.1876 & 91.1876 & 0.0000 & 0.4559 \\
        $M_W$ & 80.4468 & 80.3850 & 0.1537 & 0.4022 \\
        $1/\alpha_\text{em}$ &  137.6063 & 137.0360 & 0.8324 & 0.6852 \\
        $G_\mu\times10^5$ & 1.1739 & 1.1664 & 0.6461 & 0.0117 \\
        $\alpha_3(M_Z)$ & 0.1191 & 0.1185 & 0.7247 & 0.0008 \\
        \hline
        $M_t$ & 173.9468 & 173.2100 & 0.3468 & 2.1247 \\
        $m_b(m_b)$ & 4.3095 & 4.1800 & 0.9916 & 0.1306 \\
        $M_\tau$ & 1.7745 & 1.7769 & 0.1207 & 0.0199 \\
        \hline
        $M_b-M_c$ & 3.3159 & 3.4500 & 0.3578 & 0.3749 \\
        $m_c(m_c)$ & 1.2408 & 1.2750 & 1.1894 & 0.0288 \\
        $m_s(2\text{GeV})$ & 0.0899 & 0.0950 & 0.9886 & 0.0051 \\
        $m_d/m_s(2\text{GeV})$  & 0.0716 & 0.0513 & 3.0436 & 0.0067 \\
        $Q$ & 25.9397 & 23.0000 & 1.2742 & 2.3071 \\
        $M_\mu$ & 0.1048 & 0.1057 & 0.4139 & 0.0022 \\
        $M_e\times10^4$ & 5.1422 & 5.1100 & 0.5634 & 0.0571 \\
        \hline
        $|V_{ud}|$ & 0.9745 & 0.9742 & 0.0508 & 0.0049 \\
        $|V_{us}|$ & 0.2244 & 0.2253 & 0.6695 & 0.0014 \\
        $|V_{ub}|\times10^3$ & 3.1730 & 3.8500 & 0.7838 & 0.8637 \\
        $|V_{cd}|$ & 0.2242 & 0.2250 & 0.0967 & 0.0081 \\
        $|V_{cs}|$ & 0.9737 & 0.9860 & 0.7365 & 0.0167 \\
        $|V_{cb}|\times10^3$ & 41.1696 & 40.8000 & 0.1634 & 2.2622 \\
        $|V_{td}|\times10^3$ & 8.9578 & 8.4000 & 0.8932 & 0.6245 \\
        $|V_{ts}|\times10^3$ & 40.3083 & 40.0000 & 0.1092 & 2.8231 \\
        $|V_{tb}|$ & 0.9991 & 1.0210 & 0.6744 & 0.0324 \\
        $\sin2\beta$ & 0.6295 & 0.6820 & 2.7223 & 0.0193 \\
        $\epsilon_K$ & 0.0019 & 0.0022 & 1.2750 & 0.0002 \\
        \hline
        $\Delta M_{B_s}/\Delta M_{B_d}$ & 29.7669 & 34.8272 & 0.7181 & 7.0465 \\
        $\Delta M_{B_d}\times10^{13}$ & 3.6386 & 3.3560 & 0.4203 & 0.6723 \\
        \hline
        $m^2_{21}\times10^5$ &   7.4085 & 7.3750 & 0.0826 & 0.4057 \\
        $m^2_{31}\times10^3$ &  2.4628 & 2.5000 & 0.2818 & 0.1318 \\
        $\sin^2\theta_{12}$ & 0.2940 & 0.2975 & 0.2107 & 0.0166 \\
        $\sin^2\theta_{23}$ & 0.4573 & 0.4435 & 0.5195 & 0.0266 \\
        $\sin^2\theta_{13}$ & 0.0206 & 0.0215 & 0.8706 & 0.0010 \\
        \hline
        $M_h$ & 125.7055 & 125.0900 & 0.1279 & 4.8104 \\
        \hline
        $BR(B\to s\gamma)\times10^4$ & 3.2107 & 3.4300 & 0.1338 & 1.6389 \\
        $BR(B_s\to\mu^+\mu^-)\times10^9$ & 4.6979 & 2.8500 & 2.0300 & 0.9103 \\
        $BR(B_d\to\mu^+\mu^-)\times10^{10}$ & 1.5821 & 4.0000 & 1.3977 & 1.7299 \\
        $BR(B\to\tau\nu)\times10^5$ & 6.0704 & 11.4000 & 0.8669 & 6.1480 \\
        \hline
        $BR(B\to K^*\mu^+\mu^-)_{1\leq q^2\leq6\text{GeV}^2}\times10^8$ & 4.6258 & 3.4000 & 0.3384 & 3.6226 \\
        $BR(B\to K^*\mu^+\mu^-)_{14.18\leq q^2\leq 16\text{GeV}^2}\times10^8$ & 6.6955 & 5.6000 & 0.1025 & 10.6892 \\
        $q_0^2(A_\text{FB}(B\to K^*\mu^+\mu^-))$ & 3.8434 & 4.9000 & 0.6894 & 1.5327 \\
        $F_L(B\to K^*\mu^+\mu^-)_{1\leq q^2\leq6\text{GeV}^2}$ & 0.7524 & 0.6500 & 0.3345 & 0.3061 \\
        $F_L(B\to K^*\mu^+\mu^-)_{14.18\leq q^2\leq16\text{GeV}^2}$ & 0.3514 & 0.3300 & 0.0767 & 0.2789 \\
        $P_2(B\to K^*\mu^+\mu^-)_{1\leq q^2\leq6\text{GeV}^2}$ & 0.0660 & 0.3300 & 0.3969 & 0.6651 \\
        $P_2(B\to K^*\mu^+\mu^-)_{14.18\leq q^2\leq16\text{GeV}^2}$ & -0.4334 & -0.5000 & 0.2933 & 0.2270 \\
        $P_4'(B\to K^*\mu^+\mu^-)_{1\leq q^2\leq6\text{GeV}^2}$ & 0.5807 & 0.5800 & 0.0018 & 0.4009 \\
        $P_4'(B\to K^*\mu^+\mu^-)_{14.18\leq q^2\leq16\text{GeV}^2}$ & 1.2176 & -0.1800 & 1.9886 & 0.7028 \\
        $P_5'(B\to K^*\mu^+\mu^-)_{1\leq q^2\leq6\text{GeV}^2}$ & -0.3244 & 0.2100 & 2.3130 & 0.2311 \\
        $P_5'(B\to K^*\mu^+\mu^-)_{14.18\leq q^2\leq16\text{GeV}^2}$ & -0.7122 & -0.7900 & 0.1426 & 0.5455 \\
        \hline
        \multicolumn{3}{|l|}{Total $\chi^2$} & 47.6151 & \\
        \hline
      \end{tabular}
    }
  \end{center}
\end{table}

\section{Averaging Inflaton Oscillation}
\label{app:ave}
In this section, we calculate the average decay rates over one oscillation period, with the assumption that the inflaton oscillation is sinusoidal with period $2\pi$ and amplitude $\phi_\text{amp}$:
\begin{align}
  \phi(t) = \phi_\text{max}\sin m_\phi t \,.
\end{align}
Since the oscillation is symmetrical we only need to consider the first quarter of the oscillation, i.e.~from $m_\phi t=0$ to $m_\phi t=\pi/2$.

\subsection{Perturbative decay of the inflaton}
The decay of the inflaton to the Higgses in eq.~(\ref{eq:phi2h}) is
\begin{align}
  \Gamma_{\phi\to h} \propto \frac{\phi^2}{m_\phi}\Theta(m_\phi-2\alpha\phi) \,.
\end{align}
Due to the Heaviside Theta function, this decay rate is non-zero only when $\phi<m_\phi/2\alpha$, hence the decay only occurs from $m_\phi t=0$ to
\begin{align}
  m_\phi t_c = \sin^{-1}\frac{m_\phi}{2\alpha\phi_\text{max}} \,.
\end{align}
Hence, the average decay rate is
\begin{align}
  \langle \Gamma_{\phi\to h} \rangle
  = \frac{2}{\pi}\int_0^{m_\phi t_c}\mathrm{d}(m_\phi t)\,\frac{\phi_\text{max}^2}{m_\phi}\sin^2m_\phi t
  = \frac{2\phi_\text{max}^2}{m_\phi\pi}\left(\frac{m_\phi t_c}{2}-\frac{1}{4}\sin(2m_\phi t_c)\right) \,.
\end{align}

The decay of the inflaton to the Higgsinos in eq.~(\ref{eq:phi2fh}) is
\begin{align}
  \Gamma_{\phi\to\tilde{h}} \propto m_\phi\Theta(m_\phi-2\alpha\phi) \,.
\end{align}
Similar to the decay of Higgses, the decay only occurs from $m_\phi t=0$ to $m_\phi t=m_\phi t_c$.  Hence, the average decay rate is
\begin{align}
  \langle\Gamma_{\phi\to\tilde{h}}\rangle
  = \frac{2}{\pi}\int_0^{m_\phi t_c}\mathrm{d}(m_\phi t)\,m_\phi
  = \frac{2m_\phi^2 t_c}{\pi} \,.
\end{align}

\subsection{Decay of the Higgses}
The decays of the up-type Higgses to the right-handed neutrinos given in eq.~(\ref{eq:hu2n}), the down-type Higgses to the right-handed sneutrinos in eq.~(\ref{eq:hd2sn}), and the up-type Higgsinos to the right-handed (s)neutrinos in eq.~(\ref{eq:fhu2n}) and (\ref{eq:fhu2sn}) have the following form
\begin{align}
  \{\Gamma_{h_u\to\bar{\nu}_i^\dagger},
    \Gamma_{h_d\to\tilde{\bar{\nu}}_i},
    \Gamma_{\tilde{h_u}\to\bar{\nu}_i^\dagger},
    \Gamma_{\tilde{h_u}\to\tilde{\bar{\nu}}_i^\dagger}\}
  \propto \phi\Theta(\alpha\phi-M_{R_i}) \,.
\end{align}
These decay rates are non-zero only from
\begin{align}
  m_\phi t_c = \sin^{-1}\frac{M_{R_i}}{\alpha\phi_\text{max}} \,,
\end{align}
to $m_\phi t=\pi/2$.
Hence, the average decay rates are (for $\alpha \ \phi_{max} \geq M_{R_i}$)
\begin{align}
  \langle\{\Gamma_{h_u\to\bar{\nu}_i^\dagger},
    \Gamma_{h_d\to\tilde{\bar{\nu}}_i},
    \Gamma_{\tilde{h_u}\to\bar{\nu}_i^\dagger},
    \Gamma_{\tilde{h_u}\to\tilde{\bar{\nu}}_i^\dagger}\}\rangle
  \propto \frac{2}{\pi}\int_{m_\phi t_c}^{\pi/2}\mathrm{d}(m_\phi t)\,\phi_\text{max}\sin m_\phi t
  = \frac{2}{\pi}\phi_\text{max}\sqrt{1-\frac{M_{R_i}^2}{\alpha^2\phi_\text{max}^2}} \,.
\end{align}

The decay of the up-type Higgses to right-handed sneutrinos in eq.~(\ref{eq:hu2sn}) is
\begin{align}
  \Gamma_{h_u\to\tilde{\bar{\nu}}_i^\dagger}
  \propto \frac{M_{R_i}^2}{\phi}\Theta(\alpha\phi-M_{R_i}) \,.
\end{align}
Similar to above, this decay rate is non-zero only from $m_\phi t=m_\phi t_c$ to $m_\phi t=\pi/2$.  Hence, the average decay rate is
\begin{align}
  \langle\Gamma_{h_u\to\tilde{\bar{\nu}}_i^\dagger}\rangle
  \propto \frac{2}{\pi}\int_{m_\phi t_c}^{\pi/2}\mathrm{d}(m_\phi t)\,\frac{M_{R_i}^2}{\phi_\text{max}\sin m_\phi t}
  = -\frac{2}{\pi}\frac{M_{R_i}^2}{\phi_\text{max}}\text{ln}\tan\frac{m_\phi t_c}{2} \,.
\end{align}

On the other hand, the decays of the Higgses to radiation in eq.~(\ref{eq:hu2r}), (\ref{eq:hd2r}), (\ref{eq:fhu2r}), and (\ref{eq:fhd2r}) have the following form:
\begin{align}
  \Gamma_{\{h_u,h_d,\tilde{h}_u,\tilde{h}_d\to R}
  \propto \phi \,.
\end{align}
Hence, the average decay rates are just
\begin{align}
  \langle\Gamma_{\{h_u,h_d,\tilde{h}_u,\tilde{h}_d\to R}\rangle
  \propto \frac{2}{\pi}\phi_\text{max} \,.
\end{align}

\subsection{Decay of right-handed (s)neutrinos}
The decays of the right-handed (s)neutrinos to Higgses in eq.~(\ref{eq:n2hu}), \ref{eq:n2fhu}), and (\ref{eq:sn2hu}) have the following form
\begin{align}
  \{\Gamma_{\bar{\nu}_i\to h_u^\dagger},
    \Gamma_{\bar{\nu}_i\to\tilde{h}_u^\dagger},
    \Gamma_{\tilde{\bar{\nu}}_i\to h_u^\dagger}\}
  \propto M_{R_i}\Theta(M_{R_i}-\alpha\phi) \,.
\end{align}
The decay only occurs between $m_\phi t=0$ and
\begin{align}
  m_\phi t = m_\phi t_c = \sin^{-1}\frac{M_{R_i}}{\alpha\phi_\text{max}} \,.
\end{align}
Hence, the average decay rates are
\begin{align}
  \langle\{\Gamma_{\bar{\nu}_i\to h_u^\dagger},
    \Gamma_{\bar{\nu}_i\to\tilde{h}_u^\dagger},
    \Gamma_{\tilde{\bar{\nu}}_i\to h_u^\dagger}\}\rangle
  \propto \frac{2}{\pi}\int_0^{m_\phi t_c}\mathrm{d}(m_\phi t)\,M_{R_i}
  = \frac{2m_\phi t_c}{\pi}M_{R_i} \,.
\end{align}

The decay of the right-handed sneutrinos to the down-type Higgses in eq.~(\ref{eq:sn2hd}) is
\begin{align}
  \Gamma_{\tilde{\bar{\nu}}_i\to h_d}
  \propto \frac{\phi^2}{M_{R_i}}\Theta(M_{R_i}-\alpha\phi) \,.
\end{align}
Similar to above, the decay only occurs between $m_\phi t=0$ and $m_\phi t=m_\phi t_c$.  Hence, the average decay rate is
\begin{align}
  \langle\Gamma_{\tilde{\bar{\nu}}_i\to h_d}\rangle
  \propto \frac{2}{\pi}\int_0^{m_\phi t_c}\mathrm{d}(m_\phi t)\,\frac{\phi_\text{max}^2}{M_{R_i}}\sin^2m_\phi t
  = \frac{2\phi_\text{max}^2}{\pi M_{R_i}}\left[\frac{m_\phi t_c}{2}-\frac{1}{4}\sin(2m_\phi t_c)\right] \,.
\end{align}

\clearpage
\newpage

\bibliographystyle{utphys}
\bibliography{bibliography}

\end{document}